\documentclass[preprint]{aastex61}
\usepackage{textcomp}
\usepackage{amsmath}
\usepackage{chngcntr}

\newcommand\msun{$M_\odot$}

\received{xxx}
\revised{xxx}
\accepted{xxx}
\submitjournal{ApJ}

\usepackage{verbatim}
\shorttitle{SN\,2014C Silicate Dust}
\shortauthors{Tinyanont et al.}

\begin{document}

\title{Supernova 2014C: ongoing interaction with extended circumstellar material with silicate dust}

\correspondingauthor{Samaporn Tinyanont}
\email{st@astro.caltech.edu}

\author[0000-0002-1481-4676]{Samaporn Tinyanont}
\affiliation{Division of Physics, Mathematics and Astronomy, California Institute of Technology, 1200 E. California Blvd., Pasadena, CA 91125, USA}
\nocollaboration

\author{Ryan M Lau}
\affiliation{Institute of Space \& Astronautical Science, Japan Aerospace Exploration Agency, 3-1-1 Yoshinodai, Chuo-ku, Sagamihara, Kanagawa 252-5210, Japan}

\author{Mansi M Kasliwal}
\affiliation{Division of Physics, Mathematics and Astronomy, California Institute of Technology, 1200 E. California Blvd., Pasadena, CA 91125, USA}

\author[0000-0003-2611-7269]{Keiichi Maeda}
\affiliation{Department of Astronomy, Kyoto University, Kitashirakawa-Oiwake-cho, Sakyo-ku, Kyoto 606-8502, Japan}

\author{Nathan Smith}
\affiliation{Steward Observatory, 933 North Cherry Avenue, Tucson, AZ 85721, USA}

\author{Ori D Fox}
\affiliation{Space Telescope Science Institute, 3700 San Martin Dr, Baltimore, MD 21218, USA}

\author{Robert D Gehrz}
\affiliation{Minnesota Institute for Astrophysics, School of Physics and Astronomy, University of Minnesota, 116 Church Street, S. E., Minneapolis, MN 55455, USA}

\author{Kishalay De}
\affiliation{Division of Physics, Mathematics and Astronomy, California Institute of Technology, 1200 E. California Blvd., Pasadena, CA 91125, USA}

\author{Jacob Jencson}
\affiliation{Division of Physics, Mathematics and Astronomy, California Institute of Technology, 1200 E. California Blvd., Pasadena, CA 91125, USA}

\author{John Bally}
\affiliation{Center for Astrophysics and Space Astronomy, Department of Astrophysical and Planetary Sciences, University of Colorado, Boulder, CO 80389, USA}

\author{Frank Masci}
\affiliation{IPAC, California Institute of Technology, 1200 E. California Blvd., Pasadena, CA 91125, USA}




\begin{abstract}
Supernova (SN) 2014C is a unique explosion where a seemingly typical hydrogen-poor stripped envelope SN started to interact with a dense, hydrogen-rich circumstellar medium (CSM) a few months after the explosion.
The delayed interaction suggests a detached CSM shell, unlike in a typical SN IIn where the CSM is much closer and the interaction commences earlier post-explosion; indicating a different mass loss history.
We present near- to mid-infrared observations of SN\,2014C from 1-5 years after the explosion, including uncommon 9.7 $\mu$m imaging with COMICS on the Subaru telescope. 
Spectroscopy shows that the interaction is still ongoing, with the intermediate-width He~\textsc{I}~1.083 $\mu$m emission present out to our latest epoch 1639 days post-explosion.
The last \textit{Spitzer}/IRAC photometry at 1920 days post-explosion further confirms ongoing CSM interaction. 
The 1-10 $\mu$m spectral energy distributions (SEDs) can be explained by a dust model with a mixture of 69\% carbonaceous and 31\% silicate dust, pointing to a chemically inhomogeneous CSM. 
The inference of silicate dust is the first among interacting SNe.
An SED model with purely carbonaceous CSM dust is possible, but would require more than 0.22 \msun{} of dust, which is an order of magnitude larger than what observed in any other SNe, measured in the same way, at this epoch. 
The light curve beyond 500 days is well fit by an interaction model with a wind-driven CSM and a mass loss rate of $\sim 10^{-3} \, M_{\odot}\,\rm yr^{-1}$, which presents an additional CSM density component exterior to the constant density shell reported previously in the literature. 
SN\,2014C could originate in a binary system, similar to RY Scuti, which would explain the observed chemical and density profile inhomogeneity in the CSM.  

\end{abstract}




\section{Introduction}\label{sec:introduction}
\defcitealias{tinyanont2016}{T16}
Massive stars significantly drive the chemical evolution of the interstellar medium (ISM), with mass loss throughout their lives leading up to the final core-collapse supernova (CCSN).
However, much is unknown about the last stages of their evolution.
While we understand several physical processes behind mass loss in massive stars, such as various types of stellar winds and eruptive outflows, quantitative modeling of stellar evolution to determine the endpoint remains difficult due to uncertainties in the model assumptions.
As a result, we currently do not have a definitive model that maps classes of progenitor stars with different amount of hydrogen in their envelopes and different circumstellar media (CSM) to the resulting subtypes of CCSNe with different strengths of hydrogen features in their spectra. 
In recent years, binary interactions and episodic mass loss have become more preferred over the steady wind-driven outflow as the mass loss mechanism for massive stars as the rate expected from wind-driven mass loss is insufficient, in most cases, to strip the entire hydrogen envelope to produce progenitors to SNe Ib/c \cite[e.g.][]{demarco2017, smith2014}.
Wind-driven mass loss also cannot explain the large amount of CSM observed in some SNe that have signatures of strong CSM interactions, those of Type IIn. 
Observations of the interaction between the SN shock and the CSM allow us to probe the mass loss history, providing clues to the identity of the progenitor star. 

The CSM around some Galactic massive stars have been directly observed, showing a vast diversity in terms of density and symmetry ranging from a diffuse media around red supergiants (RSGs) like Betelgeuse \citep[e.g.][]{kervella2016, smith2009} to a massive, dense, and asymmetric nebulae around the peculiar luminous blue variable (LBV) $\eta$ Car \citep[e.g.][]{smith2006}.
One feature these nebulae have in common, however, is that they show strong signatures of silicate dust at 10-20 $\mu$m.
Some examples include the nebula around a typical LBV, HR Car \citep{umana2009}, and most RSG observed at these wavelengths \citep[e.g.][]{woods2011}.
Dust grains typically form in the cooling outflow after the formation of carbon monoxide (CO), which equally depletes C and O atoms from the medium. 
If the outflow is oxygen-rich, there are leftover O that can form precursor species to silicate dust, like $\rm SiO_2$ and $\rm Mg_2 Si O_4$; otherwise, carbonaceous dust grains can form \citep[which happens in SN ejecta as described by ][]{sarangi2013, sarangi2015}.
The presence of silicates in most Galactic massive stars indicates that their outflows are oxygen rich.
The Galactic massive stars with carbonaceous dust grains in their nebulae are primarily the carbon-rich Wolf-Rayet (H-poor) stars (WC) such as Ve-245, which produce dust when their high velocity, carbon-rich, wind decelerate as it interacts with the wind from its companion star \citep[e.g.][]{gehrz1974, marchenko2017}. 
If these WCs were to explode as a SN, it would be of Types Ib/c with large ejected mass. 
As a result, both silicate and carbonaceous dust are not expected to form in the same medium with a homogeneous C/O ratio since either C or O will be depleted first in the CO formation.

The CSM around an extragalactic massive star can be probed when the star explodes as a SN, sending out a shock wave that will interact with its surroundings. 
The shock-CSM interaction accelerates electrons to relativistic speeds, causing them to emit at X-ray and radio wavelengths. 
The X-ray photons can photoionize the unshocked gas in the CSM, which then recombines. 
In the densest CSM, the narrow emission lines from this process can be observed, constituting the Type IIn SNe, which accounts for $8.8 \substack{+3.3 \\ -2.9}\%$ of CCSNe \citep{smith2011a}. 
Pre-existing dust in the CSM gets radiatively or collisionally heated by the shock, and emit strongly in the infrared (IR). 
Population studies of SNe IIn using \textit{Spitzer Space Telescope} \citep{werner2004,gehrz2007} have shown that these SNe are luminous and long-lasting in the IR \citep[e.g.][]{fox2011,fox2013}.
As a result, IR spectral energy distribution (SED) or spectra, when available, can reveal the dust composition of the CSM, allowing for a comparison with Galactic massive stars to identify the type of the progenitor star. 
The most distinguishing features between the carbonaceous and silicate dust are the broad mid-IR silicate features at around 10 and 18 $\mu$m. 

IR observations of SNe at these long wavelengths are difficult to obtain from the ground owing to the high thermal background from the atmosphere and the telescope, and the atmospheric absorptions from water vapor. 
Hence, these observations are rare: for strongly interacting SNe IIn, only four have been observed beyond 5 $\mu$m.
Three of which, SNe 1995N \citep{vandyk2013}, 2005ip \citep{fox2010}, and 2006jd \citep{stritzinger2012} were detected by \textit{Spitzer} and/or the \textit{Wide-field InfraRed Explorer (WISE)}, with SN\,2005ip being the only one with spectroscopy.
SN\,2010jl was observed with the Faint Object infraRed CAmera for the SOFIA Telescope (FORCAST, \citealp{herter2018}) on board the Stratospheric Observatory for Infrared Astronomy \citep[(SOFIA)][]{young2012} at 11.1 $\mu$m with a deep upper limit.
In all cases, the SED could be explained by purely carbonaceous grain models, which is unexpected given that all the Galactic stars that are potentially progenitors to these strongly interacting, H-rich SNe (e.g. LBVs, RSGs mentioned above) have silicate dust in their CSM. 
We note that these four SNe may not show silicate features because the dust is optically thick, and not because there are no silicate grains.
However, the opacity effect is unlikely the explanation for SN\,1995N, which still did not show any obvious silicate features more than ten years post-explosion (see dust opacity evolution in \citealp{dwek2019}).
Given the small number of these observations due to the limited options for mid-IR observations after the \textit{Spitzer} cryogenic mission, an addition to this sample would greatly improve our understanding of the landscape of SNe emission beyond 5 $\mu$m. 

SN\,2014C was the latest nearby interacting SN for which ground-based IR observations beyond 2.5 $\mu$m were possible. 
It was first detected on 2014 January 5 (UT dates are used throughout the paper) in NGC\,7331 (Cepheid distance of 14.7 Mpc, \citealp{freedman2001}), and was quickly identified spectroscopically as a typical, non-interacting H-poor Type Ib SN \citep{kim2014}. 
The SN reached maximum light on 2014 January 13 \citep{milisavljevic2015}, which we adopted as a reference epoch throughout this paper.
The SN returned from behind the Sun and was observed on 2014 May 6, 113 days post-maximum, with an emerging H$\alpha$ emission line with an intermediate width of $\approx 1200\,\rm km\,s^{-1}$, indicating that the SN shock had started to interact with a hydrogen-rich CSM, likely the progenitor's ejected envelope \citep{milisavljevic2015}.
Such a transformation from a H-poor SN Ib/c to a H-rich interacting SN IIn had been observed before in SN\,2001em \citep{chugai2006}, and since in SN\,2004dk \citep{mauerhan2018}. 
However, SN\,2014C was the first event for which the transition was caught in action in great details.
\cite{margutti2017} reported early-time optical, ultraviolet (UV), and X-ray observations.
From the semi-bolometric luminosity obtained from the near-UV to optical photometry at early time, they derived the explosion energy and the ejected mass of $1.8 \times 10^{51} \, \rm erg$ and 1.7 \msun{}, typical for a normal SN Ib \citep{lyman2016, drout2011}. 
The X-ray flux, indicating the interaction strength, continued to rise, peaking at $\approx 400$ days, which put the location of the dense shell at $6\times 10^{16} \, \rm cm$ from the star, following the shock radius evolution calculated following \citet{chevalier1982} and \citet{chevalier1989}. 
They  inferred the total mass of the ionized CSM from the observed bremsstrahlung absorption and arrived at $\approx$ 1 \msun{}, assuming that the CSM was a shell with a thickness of $10^{16} \, \rm cm$.
\cite{anderson2017} reported radio observations of SN\,2014C at 15.7 GHz, revealing two distinct interaction phases with the transition happening between 100 and 200 days. 
The two phases corresponded to the phases before and after the Type IIn transition happened in the optical. 
All published results pointed to the progenitor of SN\,2014C being a stripped-envelope star exploding in a low density bubble as a typical SN Ib, whose forward shock began to interact strongly with its ejected, H-rich envelope located at $6\times 10^{16} \rm \, cm$ at 100-200 days post-explosion.

In addition to its peculiar early time behavior, SN\,2014C's interaction was also long lasting. 
\cite{mauerhan2018} published late time optical spectra of SN\,2014C, showing that SN\,2014C's interaction was still ongoing at day 1327, their last epoch.
The shock front ($v\approx 10000\, \rm km\, s^{-1}$) would already be at $10^{17}$ cm at that epoch, indicating that the CSM shell was more extended than previously thought. 
With a prolonged interaction with such dense medium, one might expect the forward shock of the SN to substantially slow down, but the radio imaging of the SN using very long baseline interferometry (VLBI) showed that the velocity of the shock barely decreased during the first 1000 days \citep{bietenholz2018}.
They also showed in the resolved VLBI image at 1057 days that the CSM was likely asymmetric. 
The likely explanation for these results was that the CSM was not isotropic and some parts of the shock were able to propagate freely, allowing for the interaction to continue. 

The IR evolution of SN\,2014C up to 800 days post-explosion was discussed in our previous work, \citet[][\citetalias{tinyanont2016} hereafter]{tinyanont2016}, where we summarized the IR properties of SNe of all types observed as part of the SPitzer InfraRed Intensive Transients Survey (SPIRITS; \citealp{kasliwal2017}). 
SN\,2014C stood out in our sample of H-poor SNe Ib/c and IIb as the most luminous event, even before the interaction was detected in the optical. 
The evolution of the inferred dust mass was consistent with pre-existing CSM dust being heated either radiatively or collisionally by the shock interaction with a CSM shell with constant density.
This was because the dust mass grew as $t^2$ as the (constant velocity) shock's interacting surface area grew as $r^2 \propto t^2$. 
Based on the measured dust mass and location from fitting the \textit{Spitzer} 2-point SED, we assumed a dust-to-gas ratio of 0.01 and estimated the mass loss rate that created the CSM shell up to that epoch to be $\gtrsim 10^{-2}\, M_{\odot}\,\rm yr^{-1}$ (assuming the wind velocity of 100 $\rm km\,s^{-1}$), consistent with LBV giant eruptions and also consistent with \cite{margutti2017}. 
In this work, we focus on the evolution of SN\,2014C in the IR since then.
We describe our photometric and spectroscopic observations in \textsection\ref{sec:observation}. 
We analyze the data, focusing on the SED fitting to discern the dust composition and the light curve fitting to measure the CSM density profile, in \textsection\ref{sec:analysis}. 
We discuss the two possible dust compositions of the CSM around SN\,2014C in \textsection\ref{sec:si_dust}, and the additional wind-driven CSM component exterior to the constant density shell found in previous studies in \textsection\ref{sec:LC_model}.
The discussions and conclusions are presented in \textsection\ref{sec:discussion}. 

\section{Observations}\label{sec:observation}
\subsection{Photometry}\label{sec:photometry}
\subsubsection{Near-IR 1-2.5 $\mu$m Photometry}
We obtained 5 epochs of photometry of SN\,2014C in the 1-2.5 $\mu$m \textit{JHKs} bands using the Wide-field InfraRed Camera (WIRC; \citealp{wilson2003}, with a detector upgrade described in \citealp{tinyanont2019b}) on the 200-inch Hale Telescope at Palomar Observatory (P200 hereafter).
Not every epoch had all 3 bands and some epochs had different bands taken a few days apart. 
NGC\,7331 was small enough to fit into the field of view of WIRC, so the data were taken with dithering patterns that send the galaxy around the field of view to measure and subtract the sky background. 
Science images were dark subtracted and flattened using a flat field image obtained from median combining dithered sky images with sources masked out.
The photometric zero point was determined using $\sim$30 stars in the field of view with magnitudes from the Two Micron All-Sky Survey (2MASS; \citealp{milligan1996}, \citealp{skrutskie2006}).
The data reduction was performed using an automated IR imaging pipeline described in De et al. (2019, in prep).

\begin{figure}
    \centering
    \includegraphics[width=0.7\linewidth]{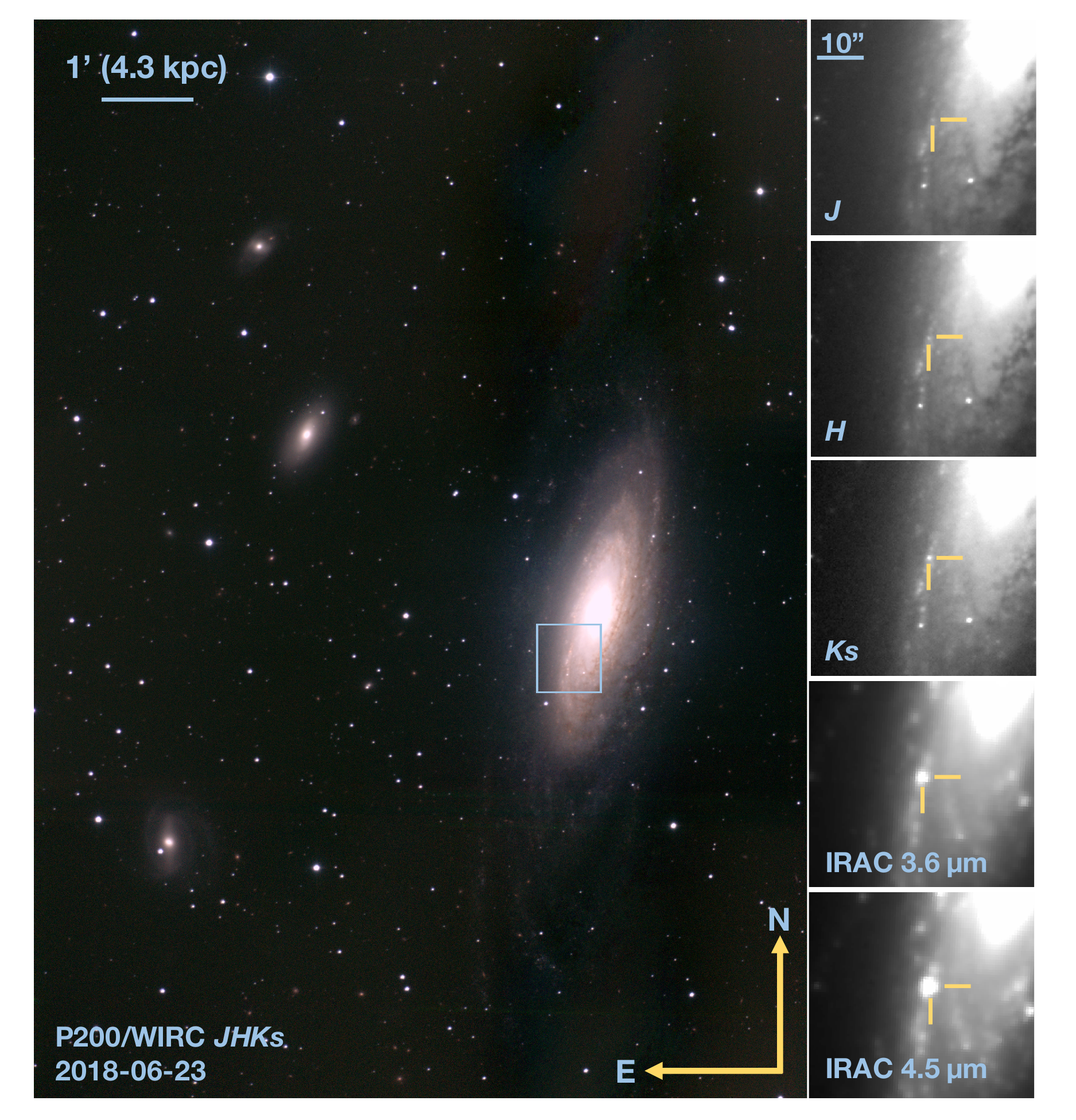}
    \caption{Left: a false image composite of NGC\,7331 where red, green, and blue correspond to \textit{Ks}, \textit{H}, and \textit{J} respectively. North is up and east is to the left. The images were taken with P200/WIRC on 2018 June 23.
    The location of the SN in the galaxy is marked with a box. 
    Right: Five images of the boxed area in the left panel in five wavelengths: \textit{J}, \textit{H}, \textit{Ks} from P200/WIRC and 3.6 and 4.5~$\mu$m from \textit{Spitzer}/IRAC. The \textit{Spitzer} observations were from 2018 February 28.}
    \label{fig:2014C_IR_image}
\end{figure}

Fig.~\ref{fig:2014C_IR_image} shows a false color image using WIRC images from 2018 June 23 where red, green, and blue correspond to $Ks$, $H$, and $J$ bands, respectively. 
Five images in the right of the figure are from the 3 near-IR bands, and the two \textit{Spitzer}/IRAC bands to visually demonstrate the red color of SN\,2014C at this epoch. 
As shown in the image, SN\,2014C was located in a spiral arm of NGC\,7331 with bright and spatially varying background, making host subtraction difficult. 
We did not have a pre-explosion image of the host galaxy at a comparable resolution to conduct image subtraction. 
As a result, we measured the near-IR photometry as follows.
First, for each band in each epoch, we created a library of point spread function (PSF) using stars in the vicinity of the SN that were not on top of the host galaxy. 
The PSF construction was done using the \texttt{EPSFBuilder} module in the \texttt{photutils} package \citep{photutils}. 
We used the \texttt{DAOStarFinder} module to find the (sub-pixel) location of the SN.
The PSF photometry was obtained by iteratively subtracting the PSF with different total flux to minimize the summed squared gradient of the image. 
This quantity was used as the metric to indicate the ``point-sourceness" of the image since the slowly varying background of the galaxy had a lower gradient than that of a point source. 
We compared the residual image of the same band from different epochs to ensure that the SN light had been consistently subtracted for all epochs. 
The PSF subtraction routine described above was custom-built in \texttt{Python}.
Along with these new photometry, we also plotted near-IR photometry published in \citetalias{tinyanont2016} in Fig.~\ref{fig:photometry}. 
All photometry discussed here, and in the following subsections, is listed in Table~\ref{tab:IR_phot}. 

\subsubsection{\textit{Spitzer}/IRAC Photometry}
SN\,2014C has been observed by the InfraRed Array Camera (IRAC; \citealp{fazio2004}) on board \textit{Spitzer} at 3.6 and 4.5~$\mu$m  in 16 epochs, as part of SPIRITS (PIDs 10136, 11063, 13053, 14089; PI Kasliwal). 
The photometry from the first 9 epochs, up to 801 days post-discovery (794 days post-maximum), were published in \citetalias{tinyanont2016}.
We had multiple pre-explosion \textit{Spitzer}/IRAC images of NGC\,7331, which we combined and used as a template for background subtraction. 
Fluxes were measured by performing aperture photometry on the background subtracted frames, and the uncertainties were estimated by combining (in quadrature) the source noise with the background noise measured from a grid of apertures placed around the source. 
We recomputed photometry for epochs that were published in \citetalias{tinyanont2016}, and found that they agree to within the uncertainty. 
We note that the updated error bars in this paper better reflect the more realistic uncertainties since the noise in subtracted images is non-Gaussian (Jencson et al. 2019, in prep). 

\subsubsection{Ground-based 3-5 $\mu$m Photometry}
SN\,2014C's brightness at 3-5 $\mu$m permitted \textit{L'} (3.43-4.13 $\mu$m) and \textit{M'} (4.55-4.79 $\mu$m) band photometry from the ground. 
We obtained those observations with the Near InfraRed Imager and spectrograph \citep[NIRI;][]{NIRI2003} on the Gemini North Telescope as part of a fast turnaround program (GN-2018A-FT-108, PI Tinyanont). 
In these thermal IR bands, the observations were background limited and we used the f/32 camera, deep-well detector bias, and fast read-out mode to maximize the observing efficiency. 
Raw images were dark subtracted and field flattened with our \texttt{python} script.
The source was observed with dithering patterns that sent the source in and out of the field of view, allowing us to subtract the bright thermal background from the sky and the telescope. 
The total on-source integration time for the \textit{L'} and \textit{M'} bands were 135.15 and 1604 seconds resulting in signal-to-noise ratios of 13 and 8, respectively.
We observed a photometric standard HD\,203856 (\textit{L'} = 6.871, \textit{M'} = 6.840; \citealp{leggett2003}) after the target for flux calibration.
The observing strategy for the standard star was similar to that of the SN, except that we only dither the star within the field of view. 
The images in the \textit{L'} and \textit{M'} bands are shown in Fig.~\ref{fig:ground_based_IR} (left and center). 
We obtained ground based 3-5 $\mu$m photometry in addition to the \textit{Spitzer}/IRAC to make sure that we have the 3-5 $\mu$m coverage concurrent with the \textit{N} band imaging described next.

\subsubsection{N-band Imaging with Subaru/COMICS}
\begin{figure}
    \centering
    \includegraphics[width = \textwidth]{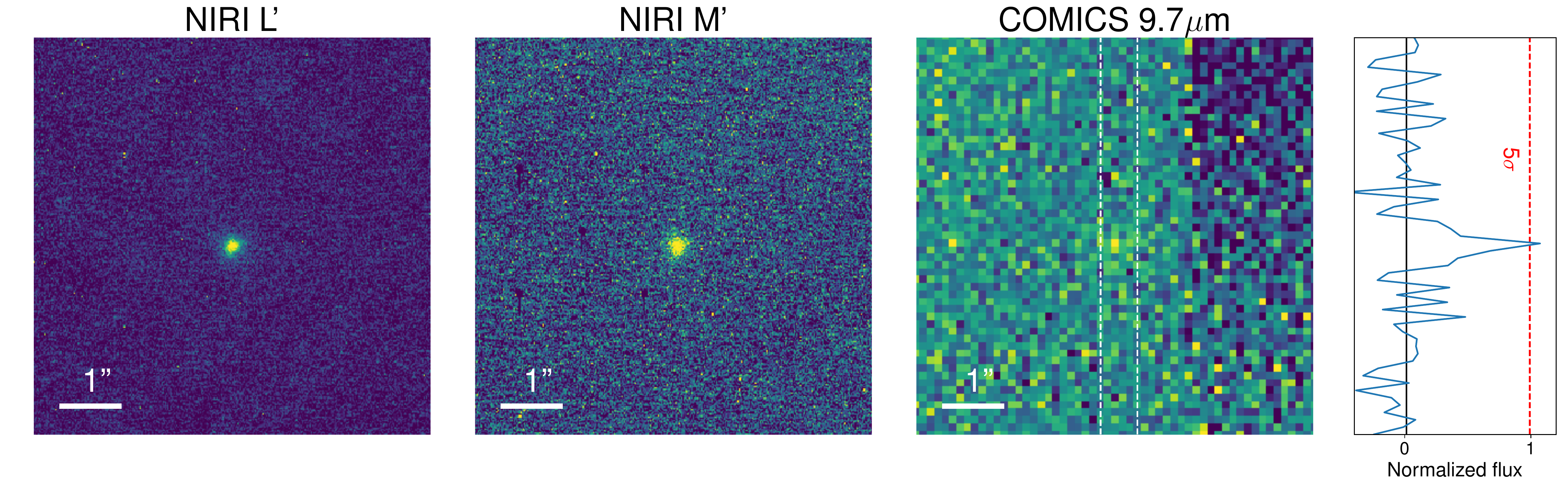}
    \caption{Ground-based IR images of SN\,2014C from NIRI in the \textit{L'} and \textit{M'} bands (left and center) and COMICS in the 9.7 $\mu$m band (right). 
    The vertical profile of the COMICS data between two dashed lines is shown in the right-most plot, with the flux normalized. The solid black line indicates the background-subtracted mean of the flux profile and the dashed red line indicates the 5$\sigma$ threshold. 
    This plot is to show that the source is detected in the COMICS image at the $5\sigma$ level.}
    \label{fig:ground_based_IR}
\end{figure}

\begin{figure}
    \centering
    \includegraphics[width = 0.8\textwidth]{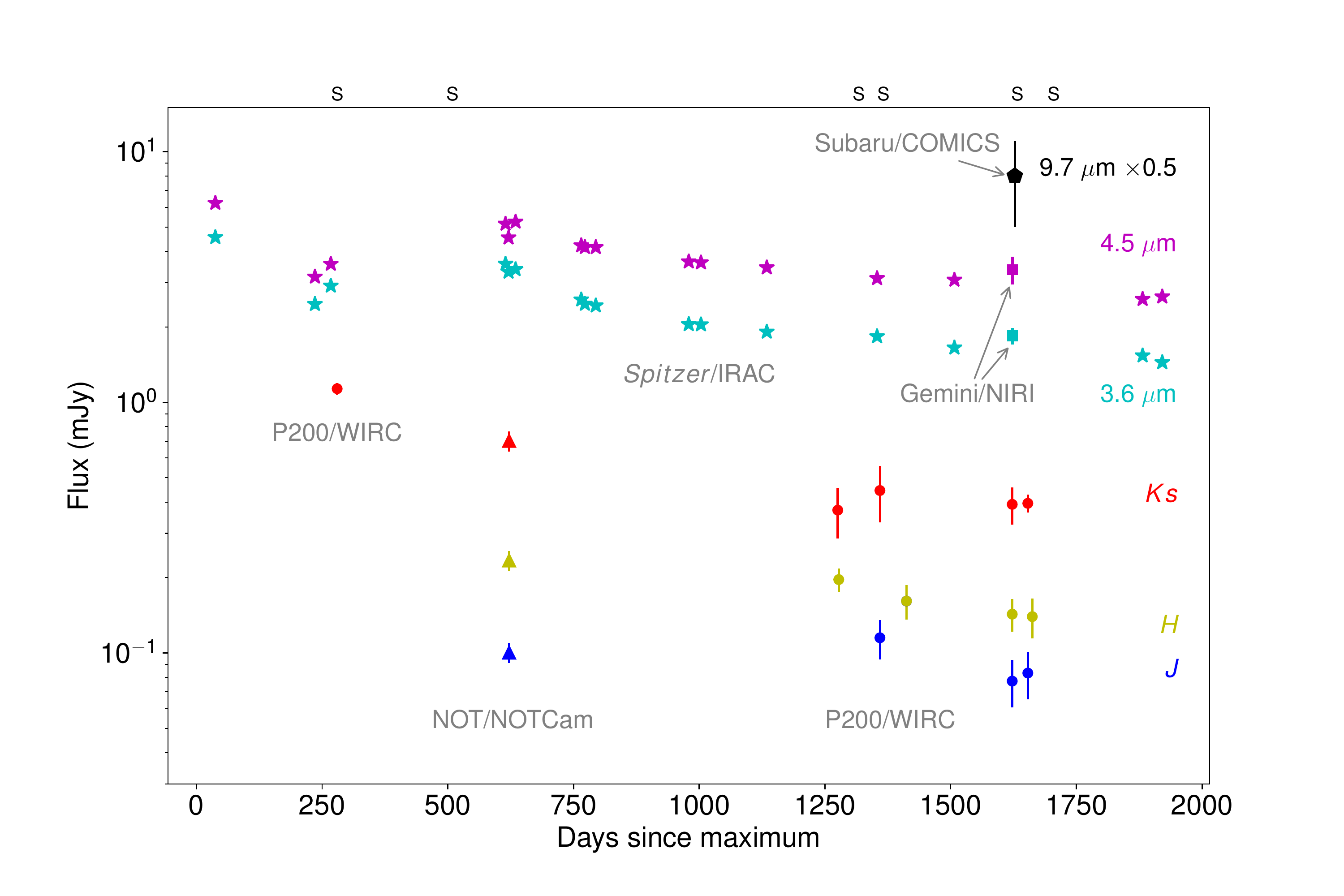}
    \caption{IR photometry of SN\,2014C in the $J$, $H$, $Ks$, 3.6 $\mu$m, 4.5 $\mu$m, and 9.7 $\mu$m bands, plotted in colors as annotated.
    Different symbols correspond to different instruments. The 9.7 $\mu$m flux is divided by two for visualization. 
    Data up to day 801 were published in \citetalias{tinyanont2016}. 
    Letters ``S" on top of the plot indicate epochs with spectroscopy.
    }
    \label{fig:photometry}
\end{figure}

SN\,2014C was observed by the Cooled Mid-Infrared Camera and Spectrometer (COMICS; \citealp{kataza2000}) on the Cassegrain focus of the Subaru Telescope with the N9.7 filter ($\lambda = 9.7$ $\mu$m, $\Delta\lambda=0.9$ $\mu$m) on 2018 June 28. 
Individual 200 s exposures were taken in chopping-only mode with a chop amplitude of 12'' and the total integration time on SN\,2014C was 1.6 hr. 
5 LAC (HR 8572) was used as the photometric standard from the list of mid-IR standards in \cite{cohen1999}. 
The COMICS N9.7 flux from 5 LAC was derived by integrating the filter bandpass over its IR spectrum provided by \cite{cohen1999}. 
Assuming a box function for the filter profile ranging from $\lambda=9.25 - 10.15$ $\mu$m, the COMICS N9.7 flux from 5 LAC was determined to be 44.0 Jy. 
We note that this is closely consistent with the calibrated flux derived with the similar VLT/VISIR B9.7 filter for the same standard (42.5 Jy\footnote{\url{http://www.eso.org/sci/facilities/paranal/instruments/visir/tools/zerop_cohen_Jy.txt}}). 
The measured full-width at half maximum (FWHM) of the calibrator was $\sim0.5''$, which was slightly worse than diffraction limited performance at 9.7 $\mu$m ($0.25''$) and likely due strong winds and poor seeing conditions during the observations.
SN\,2014C was detected with a 5-$\sigma$ significance at a flux of 16$\pm 3$ mJy in the N9.7 filter.
Fig.~\ref{fig:ground_based_IR} (right image) shows the calibrated image from COMICS. 
The rightmost plot shows the vertical profile of the flux on the source location, with a 5$\sigma$ threshold indicated to illustrate the source detection significance. 

\begin{table}
\centering
\label{tab:IR_phot}
\caption{IR photometry of SN\,2014C}
\begin{tabular}{lllllllllllll}
\toprule
Date & Epoch & $F_{J}$ & $\sigma_{F_{J}}$ & $F_{H}$ & $\sigma_{F_{H}}$ & $F_{K_s}$ & $\sigma_{F_{K_s}}$ &  $F_{[3.6]}$ & $\sigma_{F_{[3.6]}}$ &  $F_{[4.5]}$ & $\sigma_{F_{[4.5]}}$ & Instrument  \\
(UT)     & (day) & \multicolumn{10}{c}{\hrulefill \, (mJy) \hrulefill } \\
\hline
2014-02-19 & 37.8 & --- & --- & --- & --- & --- & --- & 4.55 & 0.08 & 6.24 & 0.09     & \textit{Spitzer}/IRAC  \\
2014-09-05 & 235.8 & --- & --- & --- & --- & --- & --- & 2.46 & 0.10 & 3.17 & 0.06    & \textit{Spitzer}/IRAC  \\
2014-10-07 & 267.4 & --- & --- & --- & --- & --- & --- & 2.92 & 0.09 & 3.56 & 0.06    & \textit{Spitzer}/IRAC  \\
2015-09-19 & 614.8 & --- & --- & --- & --- & --- & --- & 3.57 & 0.09 & 5.15 & 0.07    & \textit{Spitzer}/IRAC  \\
2015-09-25 & 620.8 & --- & --- & --- & --- & --- & --- & 3.31 & 0.09 & 4.54 & 0.07    & \textit{Spitzer}/IRAC  \\
2015-10-09 & 634.8 & --- & --- & --- & --- & --- & --- & 3.39 & 0.09 & 5.25 & 0.06    & \textit{Spitzer}/IRAC  \\
2016-02-17 & 765.4 & --- & --- & --- & --- & --- & --- & 2.57 & 0.01 & 4.22 & 0.02    & \textit{Spitzer}/IRAC  \\
2016-02-24 & 772.7 & --- & --- & --- & --- & --- & --- & 2.47 & 0.01 & 4.16 & 0.02    & \textit{Spitzer}/IRAC  \\
2016-03-17 & 794.3 & --- & --- & --- & --- & --- & --- & 2.43 & 0.01 & 4.15 & 0.02    & \textit{Spitzer}/IRAC  \\
2016-09-18 & 979.2 & --- & --- & --- & --- & --- & --- & 2.05 & 0.02 & 3.64 & 0.02    & \textit{Spitzer}/IRAC  \\
2016-10-12 & 1003.5 & --- & --- & --- & --- & --- & --- & 2.04 & 0.02 & 3.61 & 0.02   & \textit{Spitzer}/IRAC  \\
2017-02-20 & 1134.4 & --- & --- & --- & --- & --- & --- & 1.91 & 0.02 & 3.45 & 0.02   & \textit{Spitzer}/IRAC  \\
2017-07-11 & 1275.5 & --- & --- & --- & --- & 0.37 & 0.08 & --- & --- & --- & ---     & P200/WIRC  \\
2017-07-13 & 1277.5 & --- & --- & 0.20 & 0.02 & --- & --- & --- & --- & --- & ---     & P200/WIRC  \\
2017-09-27 & 1353.7 & --- & --- & --- & --- & --- & --- & 1.83 & 0.01 & 3.12 & 0.01   & \textit{Spitzer}/IRAC  \\
2017-10-03 & 1359.4 & 0.11 & 0.02 & --- & --- & 0.44 & 0.11 & --- & --- & --- & ---   & P200/WIRC  \\
2017-11-25 & 1412.2 & 0.16 & 0.02 & 0.16 & 0.03 & --- & --- & --- & --- & --- & ---   & P200/WIRC  \\
2018-02-28 & 1507.5 & --- & --- & --- & --- & --- & --- & 1.65 & 0.02 & 3.08 & 0.01   & \textit{Spitzer}/IRAC  \\
2018-06-23 & 1622.5 & 0.08 & 0.02 & 0.14 & 0.02 & 0.39 & 0.07 & --- & --- & --- & --- & P200/WIRC  \\
2018-06-24 & 1623.0 & --- & --- & --- & --- & --- & --- & 1.84 & 0.14 & 3.38 & 0.43 & Gemini/NIRI\tablenotemark{a} \\
2018-07-24 & 1653.5 & 0.08 & 0.02 & --- & --- & 0.40 & 0.03 & --- & --- & --- & ---   & P200/WIRC  \\
2018-08-02 & 1662.3 & --- & --- & 0.14 & 0.03 & --- & --- & --- & --- & --- & ---     & P200/WIRC  \\
2019-03-09 & 1881.8 & --- & --- & --- & --- & --- & --- & 1.54 & 0.02 & 2.58 & 0.02   & \textit{Spitzer}/IRAC  \\
2019-04-17 & 1920.8 & --- & --- & --- & --- & --- & --- & 1.44 & 0.02 & 2.64 & 0.02   & \textit{Spitzer}/IRAC  \\
\hline
2018-06-28 & 1627.5 &   &  &  & &  &  &  \multicolumn{2}{c}{$F_{[9.7]} = 16$} &\multicolumn{2}{c}{$\sigma_{F_{[9.7]}} = 3$} & Subaru/COMICS \\
\hline
\end{tabular}
\tablenotetext{a}{Gemini/NIRI's filters are \textit{L'} and \textit{M'}, which are different from \textit{Spitzer}/IRAC's filters.}
\end{table}

\subsection{Ground-based Spectroscopy}
We obtained near-IR 1-2.5 $\mu$m spectra of SN\,2014C in 7 epochs spanning from 1-5 years post-maximum.
Table~\ref{tab:spec_log} summarizes all the spectra taken. 
We used the twin medium resolution, long-slit, echelette spectrographs: TripleSpec on P200 \citep{herter2008} and the Near-Infrared Echellette Spectrometer (NIRES) on the Keck telescope\footnote{\url{https://www2.keck.hawaii.edu/inst/nires/}}, both of which provided simultaneous 1-2.5 $\mu$m spectra of a single source. 
The only difference between TripleSpec and NIRES was the fixed slit widths of 1" vs 0.55\farcs \ 
The data were taken by either dithering the source along the slit or alternating between the source and the sky to enable sky subtraction.
Data from both TripleSpec and NIRES were reduced using a version of \texttt{Spextool} \citep{cushing2004} specifically for TripleSpec and NIRES.
The software applied field flattening, retrieved a wavelength solution from sky lines present in science observations, and subtracted each exposure pair to remove most of the sky emission.
It located the trace of the object in the reduced 2D images, then fit a low order polynomial across the trace to estimate host background. 
Finally, the spectral trace was reduced using an optimal extraction algorithm to enhance the signal to noise ratio while rejecting bad pixels and cosmic ray hits \citep{horne1986}. 
The telluric and flux calibrations were performed using \texttt{xtellcor} with the standard obsevations of an A0V star either before or after the SN observation \citep{vacca2003}.

\begin{figure}
    \centering
    \includegraphics[width=\textwidth]{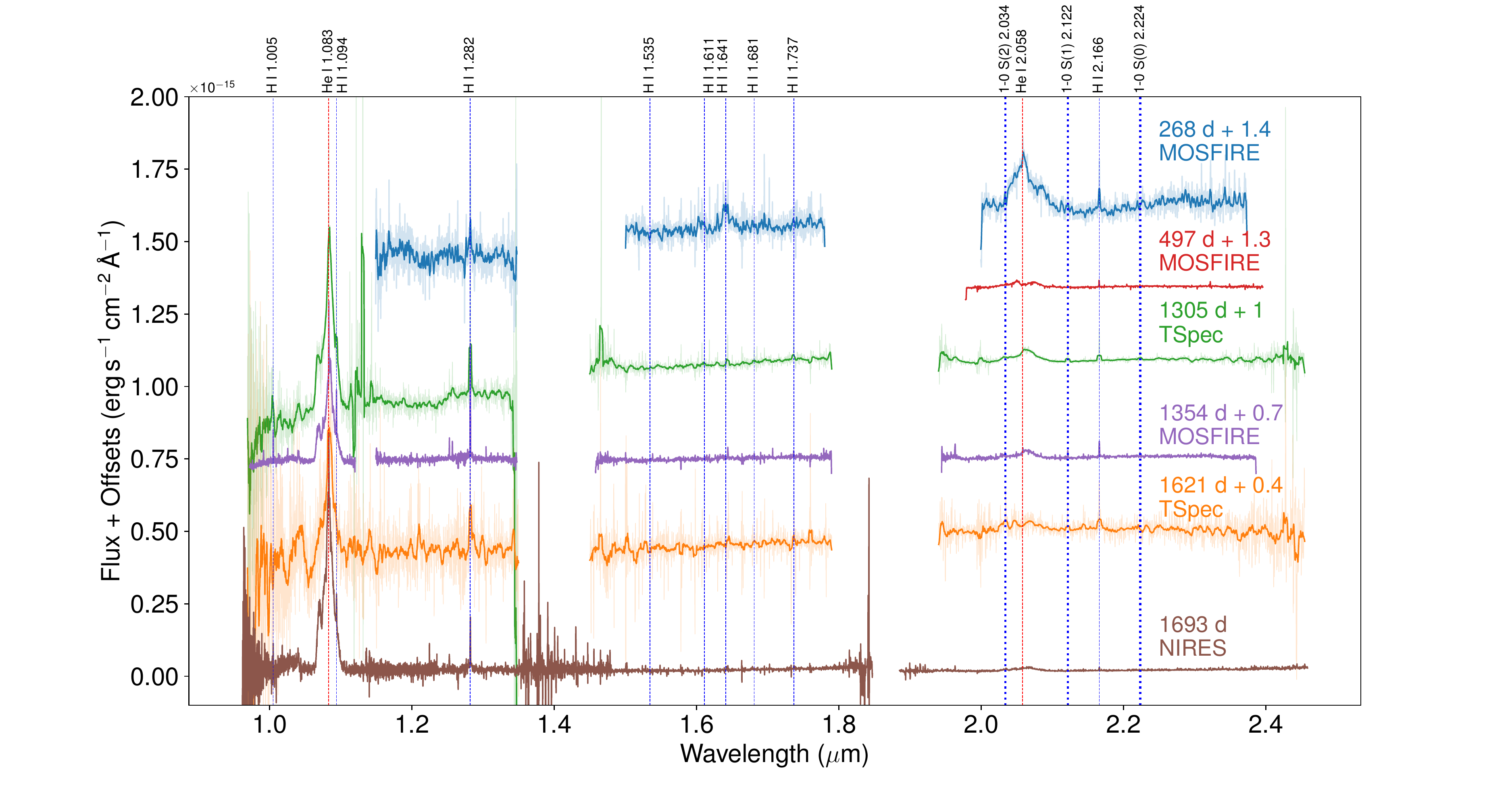}
    \caption{Near-IR spectra of SN\,2014C from 268 to 1693 days post-maximum. 
    Offsets applied to the spectra are indicated next to the corresponding dates. 
    We note that TripleSpec spectra have poorer background subtraction, in comparison to MOSFIRE or NIRES spectra, due to the higher background at Palomar and also more galaxy contamination due to the wider slit. 
    Hydrogen and helium lines, including molecular hydrogen $v=1-0$ transitions, are marked. 
    See text and Table~\ref{tab:spec_log} for details of the observations.}
    \label{fig:nir_spec}
\end{figure}

In addition, we also used the Multi-Object Spectrometer for Infra-Red Exploration \citep[MOSFIRE;][]{mclean2012} on the Keck telescope.
MOSFIRE can obtain spectra of up to 46 sources in its field of view, one filter at the time.
The filters used in each epoch are reported in Table~\ref{tab:spec_log}.
Exposure times for MOSFIRE observations are given for each filter.
The data reduction and spectral extraction were performed using MOSFIRE's data reduction pipeline\footnote{\url{https://keck-datareductionpipelines.github.io/MosfireDRP/}}, and telluric and flux calibrations were performed using \texttt{xtellcor}.
We caution that the accuracy of flux calibration between different epochs and different instrument can vary up to a factor of 2-3. 
Galaxy host contamination also varied among spectra due to the different slit widths and observing conditions.
However, all subsequent analyses of the spectra will not rely on absolute flux calibration. 
All 1-2.5 $\mu$m spectra are shown in Fig.~\ref{fig:nir_spec}. 

Finally, we obtained 3.0-3.9 $\mu$m spectroscopy using the Gemini Near-InfraRed Spectrograph (GNIRS) on the Gemini North telescope as part of a fast turnaround program (GN-2018A-FT-108, PI Tinyanont). 
Ground-based observations in this wavelength is difficult owing to the very strong thermal background from both the sky and the telescope.
For these observations, we used the long red camera with the pixel scale of 0.05"/pixel and the 0.15" slit with the 10 l/mm grating for the resolving power of $R\sim1200$. 
The wavelength range captured in our spectra was 3.1-3.9 $\mu$m. 
To handle high background, we set the detector bias to the deep well mode to increase the saturation threshold, and the read-out mode to match the read-out noise to the sky background noise.
The observations were taken on 2018 August 16-17, with the total integration time of 23 and 68 min on the respective nights.
We found, however, that only first 17 images from the first night contain the SN trace.
Each exposure was 30 sec, 2 coadds, in ABBA dithering pattern along the slit.
The data were reduced using Gemini's IRAF-based data reduction pipeline.\footnote{\url{https://www.gemini.edu/sciops/data-and-results/processing-software}}
The telluric and flux calibrations were performed using \texttt{xtellcor} with HIP\,114714 (HD\,219290) as a standard star.
We caution that spectral features present are likely spurious and the spectra simply probed the continuum. 
We overplotted the binned spectrum in Fig.~\ref{fig:IIn_comparison}, but did not use it to fit the SED. 

\begin{table}
\centering
\caption{Log of spectroscopic observations of SN\,2014C}
\label{tab:spec_log}
\begin{tabular}{lllll}
\toprule
Date & Epoch (day) & Telescope/Instrument & Bands & Exposure Time (sec)  \\
\hline
2014-10-08 & 268 &  Keck/MOSFIRE & JHK & 1440, 2160, 2160 \\
2015-05-25 & 497 &  Keck/MOSFIRE & K & 1080 \\
2017-08-09 & 1304 & P200/TripleSpec & YJHK & 5400 \\
2017-08-11 & 1306 & P200/TripleSpec & YJHK & 9600 \\
2017-09-28 & 1354 & Keck/MOSFIRE & YJHK & 720, 480, 480, 720 \\
2018-06-22 & 1621 & P200/TripleSpec & YJHK & 3900 \\
2018-08-17 & 1677 & Gemini/GNIRS & L' & 510 \\
2018-09-02 & 1693 & Keck/NIRES & YJHK & 600 \\
\hline
\end{tabular}
\end{table}

\section{Analysis}\label{sec:analysis}


\subsection{SED Fitting and Silicate Dust in the CSM}\label{sec:si_dust}
Thermal emission arises from dust grains at different temperatures and different locations in the SN. 
To roughly determine this, we first fit a simple black body to the IR SED of SN\,2014C at 1623 d, the epoch at which we have the 10 $\mu$m data and for which we will perform full SED fitting.
We find that the near-IR (1-2.5 $\mu$m) flux is well-explained by a hot component at $T\sim 1000$ K located at $r \sim 10^{15}$ cm. 
The 3-5 $\mu$m flux arises from a warm component with $T\sim 500$ K and $r \sim 4\times 10^{16}$ cm. 
Finally, the 10 $\mu$m flux can be explained by a cold component with $T\sim 300$ K and $r \sim 1.5\times 10^{17}$ cm. 
We will argue later that the 10 $\mu$m flux is more likely from silicate dust emission feature, and not a cold dust component. 
We note that the black body radii are the physical radii of optically thick dust clouds.
In reality, the CSM dust is optically thin, and the black body radii are lower limits to the actual location of the dust grains. 
The shock front at this epoch, which is determined from the model that we will discuss in \textsection\ref{sec:LC_model}, is at $r_{\rm sh} = 1.2 \times 10^{17}$ cm, suggesting that the hot component arises from likely newly-formed dust inside the ejecta while the warm component is likely associated with pre-existing CSM dust heated by the shock either radiatively or collisionally.  

We determine the composition of the dust in the CSM around SN\,2014C by fitting its IR SED with different dust models.
Following \citetalias{tinyanont2016} and \cite{fox2010}, we fit the SED with a modified black body spectrum
\begin{equation}\label{eq:duxt_flux}
    F_\nu = \dfrac{M_{\rm dust} B_{\nu}(T_{\rm dust}) \kappa_{\nu}(a)} {d^2} 
\end{equation}
where $M_{\rm dust}$ is the total dust mass, $B_\nu$ is the Planck function, $T_{\rm dust}$ is the dust temperature, and $\kappa_{\nu}(a)$ is the dust grain opacity, which is a function of dust grain size, $a$. 
This dust SED assumes that the emitting dust is optically thin, and at a single temperature.
We will check the optical depth assumption after calculating the dust mass. 
We note that this equation poses no geometric requirements on the distribution of the dust grains.
The dust emissivity ($Q$), and consequently opacity ($\kappa$) for carbonaceous and silicate grains, are from \cite{draine1984} and \cite{laor1993}, where we only consider graphite for carbonaceous grains. 
We convert the tabulated emissivity into opacity using the relation $\kappa = 3Q/4\rho_{\rm bulk} a$, where the bulk densities are $\rho_{\rm bulk} = 2.2$ and $3\, \rm g\,cm^{-3}$ for the graphite and silicate grains respectively. 
Amorphous carbon has higher emissivity, requiring less mass to explain the same IR flux, but the general shape of the SED is similar to that of graphite. 
We assume the grain size of 0.1 $\mu$m, noting that the emissivities from different grain sizes are degenerate in the IR, unless we include the unlikely large 1 $\mu$m grains (see Figure 4, top panel from \citealp{fox2010}).    

\begin{figure}
    \centering
    \includegraphics[width = \textwidth]{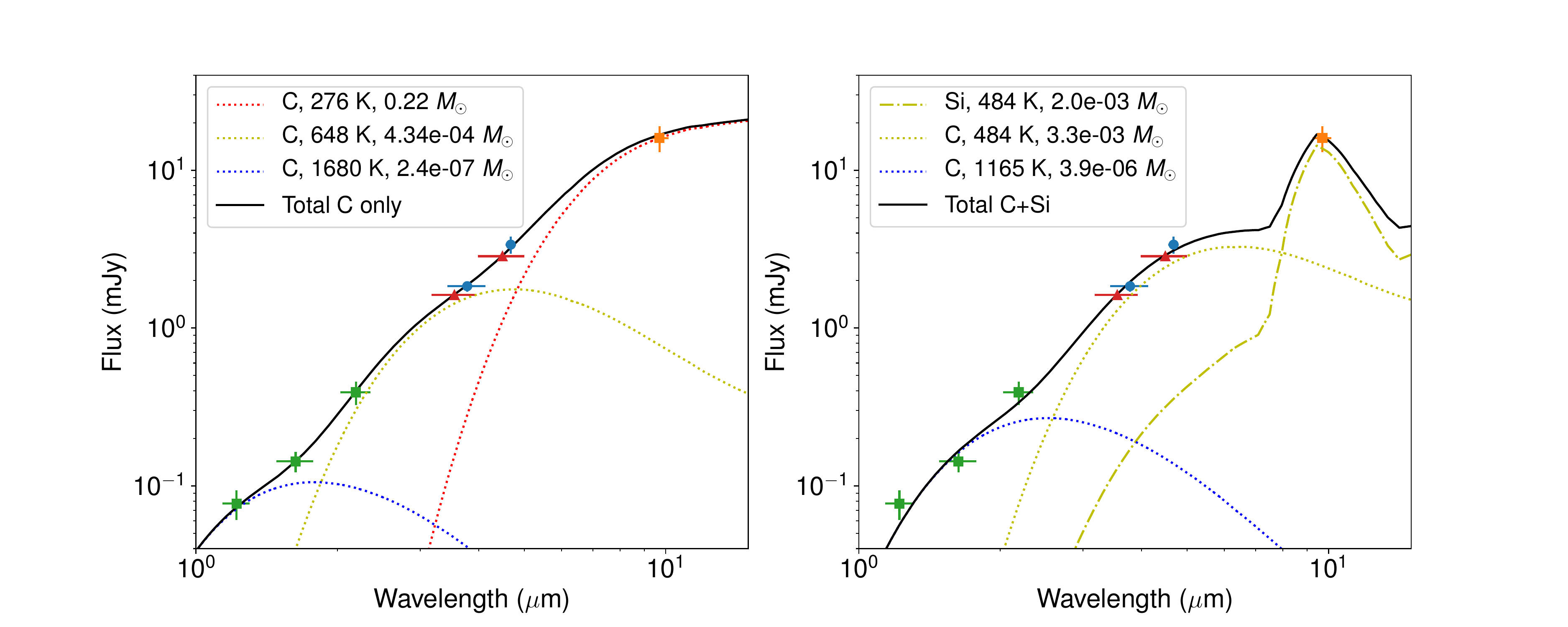}
    \caption{SED of SN\,2014C from 1 to 10 $\mu$m observed 1622-1626 days post maximum. The error bars in the \textit{x}-direction represent the filter bandwidth. The left panel shows the SED fitted by a dust emission model with only amorphous carbonaceous dust, which required three different temperatures. The right panel shows the fit with hot carbonaceous dust and warm dust with 66\% carbonaceous and 34\% silicate grains (by mass). The mass ratio was obtained from fitting the data. We note that the purely carbonaceous dust model requires an order of magnitude more dust than ever observed in previous SNe at these epochs. 
    }
    \label{fig:midIR_SED}
\end{figure}

Fig.~\ref{fig:midIR_SED} shows the SED of SN\,2014C with near-IR photometry from P200/WIRC, mid-IR photometry from \textit{Spitzer}/IRAC and Gemini/NIRI, and finally mid-IR (9.7 $\mu$m) photometry from Subaru/COMICS. 
All data were taken between 1622 and 1626 days post maximum.
The \textit{Spitzer} data were interpolated from days 1515 and 1889 using a power law. 
The two panels of Fig.~\ref{fig:midIR_SED} compare two different models, which provide different explanations for the 9.7 $\mu$m flux: copious cold carbonaceous dust (left), and the silicate dust's feature (right). 

We first consider a scenario where all CSM dust around SN\,2014C is carbonaceous (Fig.~\ref{fig:midIR_SED}, left).
This composition choice is based on prior observations showing that no other interacting SNe with mid-IR observations showed silicate features.
In this model, the 9.7 $\mu$m flux is dominated by cold carbonaceous dust at 276 K. 
We sample the parameter space with a Markov Chain Monte Carlo (MCMC) routine \texttt{emcee} \citep{emcee} to derive the model parameters along with their robust uncertainties.
More details on our MCMC sampling method are given in Appendix~\ref{sec:MCMC_SED} and the corner plot of this MCMC fitting are described in Fig.~\ref{fig:corner_SED_C}. 
While this model fits our data well, the high mass of cold carbonaceous dust necessary to produce the observed 9.7 $\mu$m flux, 0.22 \msun is problematic.  
First, 0.22~\msun{} of carbonaceous dust at a minimum distance of $r_{\rm bb} = 1.5 \times 10^{17} \, \rm cm$ (the black body radius, derived earlier in this section for a cold dust component, also consistent with the shock radius derived later in \textsection\ref{sec:LC_model}) would have maximum optical depths, $\tau_\nu = M_d \kappa_\nu / 4\pi r_{\rm bb}^2$ = 3, 2, and 0.6 at 3.6, 4.5 and 9.7 $\mu$m respectively. 
This means that the dust could be not optically thin, resulting in even higher total dust mass than observed.
Secondly, this dust mass is an order of magnitude higher than that dust mass observed in any SNe at this epoch (see e.g. Fig.~4 \citealp{gall2014}) and is an order of magnitude higher than the dust mass inferred in SN\,2006jd from 3-10 $\mu$m observations at a similar epoch \citep{stritzinger2012}.
We also note that SN\,2006jd had $\lambda > 5 \mu$m observations with neither signs of silicates nor cold ($\sim$300 K) carbonaceous dust components that appear in SN\,2014C. 
As mentioned earlier, the dust mass inferred here is the mass of the IR emitting dust, which is only a lower limit to the total dust mass in the CSM. 
Assuming the canonical gas-to-dust mass ratio of 100, the lower limit of the emitting dust mass indicates that the mass of the gas shell containing the IR emitting dust is $\gtrsim$20 \msun, which is large in comparison to even the most massive Galactic LBV nebulae \citep[][these authors assumed the same gas-to-dust ratio]{kochanek2011, smith_owocki2006}. 
Furthermore, we will show in \textsection\ref{sec:LC_model} that the total CSM mass around SN\,2014C could only be about 4-9 \msun{}.
These are the major caveats to the purely carbonaceous dust scenario. 

We now consider a second scenario, in which the 9.7 $\mu$m flux of SN\,2014C is explained by the broad silicate feature around 10 $\mu$m  (Fig.~\ref{fig:midIR_SED}, right).
In this model, we fit the SED with two dust components.
The first component is a single-temperature warm dust ($T_{\rm warm}$) composed of both carbonaceous and silicate grains with a total mass of $M_{\rm warm}$, a fraction $f_{\rm Si}$ of which is silicate.
The second component is hot dust at the temperature $T_{\rm hot}$ composed of only carbonaceous dust with the total mass of $M_{\rm hot}$.
We used MCMC as described previously and in Appendix~\ref{sec:MCMC_SED} to derive the model parameters along with their robust uncertainties.
The composition of this hot component does not matter since the flux at 10 $\mu$m from this component is negligible. 
The resulting parameters from the MCMC fits are 
$T_{\rm warm}  =  484 \substack{+ 11 \\ - 13 }$ K,
$M_{\rm warm}  =  5.2 \substack{+ 0.7 \\ - 0.6 }\times 10^{-3}$ \msun{},
$T_{\rm hot}  =  1165 \substack{+ 140 \\ - 100 }$ K,
$M_{\rm hot}  =  4 \substack{+ 4\\ - 2 } \times 10^{-6}$ \msun{}, and
$f_{\rm Si}  =  0.38 \substack{+ 0.06 \\ - 0.08 } $.
The corner plot of the MCMC fitting is presented in Fig.~\ref{fig:corner_SED}. 
The uncertainties provided here only incorporate uncertainties in the measured flux, and not in the dust emissivity.
The near-IR photometry is well fitted by a hot, 1165 K, carbonaceous dust component. 
The total warm dust mass in this scenario, $5\times 10^{-3}$ \msun{}, is similar to that observed in SN\,2006jd \citep{stritzinger2012}.
The dust cloud of this mass could also be optically thin if dust is located around the shock radius of $1.2 \times 10^{17}$ cm; the maximum optical depths are 0.01, 0.01, and 0.1 at 3.6, 4.5 and 9.7 $\mu$m, respectively. 
Because of the more reasonable dust mass required to fit the SED, we favor this scenario in which the 9.7 $\mu$m is dominated by the broad silicate dust emission. 
Follow-up observations with more photometric bands around the 9.7 $\mu$m band are required to robustly identify the silicate feature.
The mixture of silicate and carbonaceous dust grains required to explain the SED suggests that the CSM is inhomogeneous in its composition since a homogeneous medium with C/O $\neq$ 1 will form only either carbonaceous or silicate dust.  
We will further discuss the implication on the origin of the CSM around SN\,2014C in \textsection\ref{sec:discussion}.

\subsection{Comparison to Other Interacting SNe with 10 $\mu$m Observation}

\begin{figure}
    \centering
    \includegraphics[width = 0.7\textwidth]{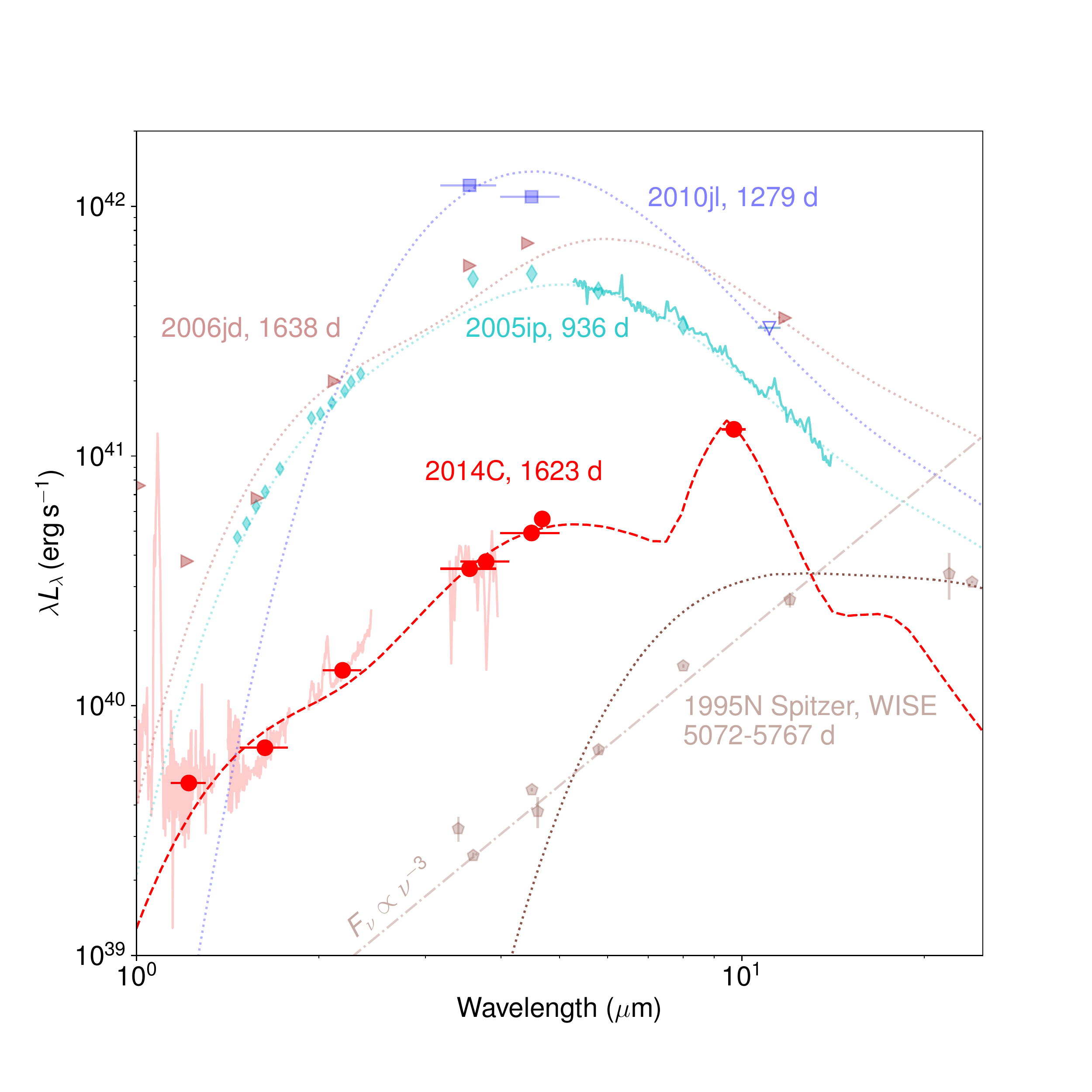}
    \caption{SED of SN\,2014C compared with all other H-rich interacting SNe with mid-IR ($> 10\, \rm \mu m$) data in the literature, namely SN\,1995N \citep{vandyk2013}, 2005ip \citep{fox2010}, 2006jd \citep{stritzinger2012}, and 2010jl \citep{williams2015}.
    Transparent red line represents the 1-2.5 $\mu$m spectra from NIRES and 3.3-3.9 $\mu$m spectrum from GNIRS taken at 1677 and 1693~d, with the flux scaled to match that from photometry. 
    Diamond points for SN\,2005ip around 1-2 $\mu$m are also taken from spectra shown in \cite{fox2010}. 
    Apart from SN\,2010jl's 11.1 $\mu$m upper limit, all other data points are detection.
    We note that the SED of SNe\,2005ip, 2006jd, and 2010jl could be explained with purely carbonaceous dust models, overplotted (dust parameters taken from respective papers).
    The SEDs of SN\,1995N, taken at a much later epoch (5000-6000 days post-explosion), were ill-fitted by either carbonaceous or silicate dust. 
    }
    \label{fig:IIn_comparison}
\end{figure}

Fig.~\ref{fig:IIn_comparison} compares the IR SED of SN\,2014C with the SEDs of all other interacting SNe for which observations beyond 5 $\mu$m are available in the literature. 
Apart from SN\,2014C, only four other strongly interacting SNe IIn have been observed in the mid-IR. 
Three of these were observed at epochs comparable to the epoch at which SN 2014C was observed in the mid-IR.
SN\,2005ip was observed at 936 days post-explosion during the cryogenic phase of \textit{Spitzer}. 
It was the only interacting SN with an InfraRed Spectrograph \citep[IRS][]{houck2004} mid-IR spectrum from 5-12 $\mu$m and IRAC photometry at 5.8 and 8 $\mu$m \citep{fox2010}. 
SN\,2006jd was observed at 1638 days, an epoch very similar to that of SN\,2014C, with \textit{Spitzer}/IRAC and \textit{WISE} \citep{stritzinger2012}.
SN\,2010jl was observed at 1279 days with \textit{Spitzer}/IRAC and SOFIA/FORCAST \citep{herter2018} at 11.1 $\mu$m, resulting in a deep upper limit after 6400 s of total integration time \citep{williams2015}. 
In all three cases, the photometry and spectra from 1-10 $\mu$m are well fitted by purely carbonaceous dust models, which we overplot in Fig~\ref{fig:IIn_comparison} using dust parameters from the literature.
Lastly, SN\,1995N was observed at more than ten years post-explosion by \textit{Spitzer}/IRAC and \textit{WISE} \citep{vandyk2013}.
Its SED shape differs markedly from those of the other three SNe observed at earlier epochs. 
The carbonaceous dust model that \cite{vandyk2013} fitted to the data is shown in Fig.~\ref{fig:IIn_comparison}.
However, we note that the shallow slope from 3-10 $\mu$m cannot be fitted with a single-temperature dust model, regardless of composition, and would require a range of dust temperatures.  
Provided the late epoch of the observation, one might also consider a non-thermal origin for the IR emission from SN\,1995N, since the SED can also be described with a broken power law ($F_{\nu} \propto \nu^{-3}$; also overplotted) with a knee at around 12 $\mu$m. 
In addition to these H-rich interacting SNe, the H-poor interacting SN\,2006jc (Ibn) was also observed beyond 5 $\mu$m with \textit{AKARI} \citep{sakon2009}.
Its SED was also best-fitted with a two-temperature amorphous carbon dust model.
This comparison highlights SN\,2014C's unique SED shape among other interacting SNe for which data are available beyond 5 $\mu$m at comparable epochs, showing for the first time an evidence for a silicate dust feature in the IR SED of an interacting SN. 
It also accentuates the need for observations of interacting SNe at late times, out to decades post-explosion, in the near- to mid-IR, which will be enabled by the upcoming \textit{James Webb Space Telescope} (\textit{JWST}). 

\subsection{Dust Parameter Evolution}
\begin{figure}
    \centering
    \includegraphics[width = 0.8\textwidth]{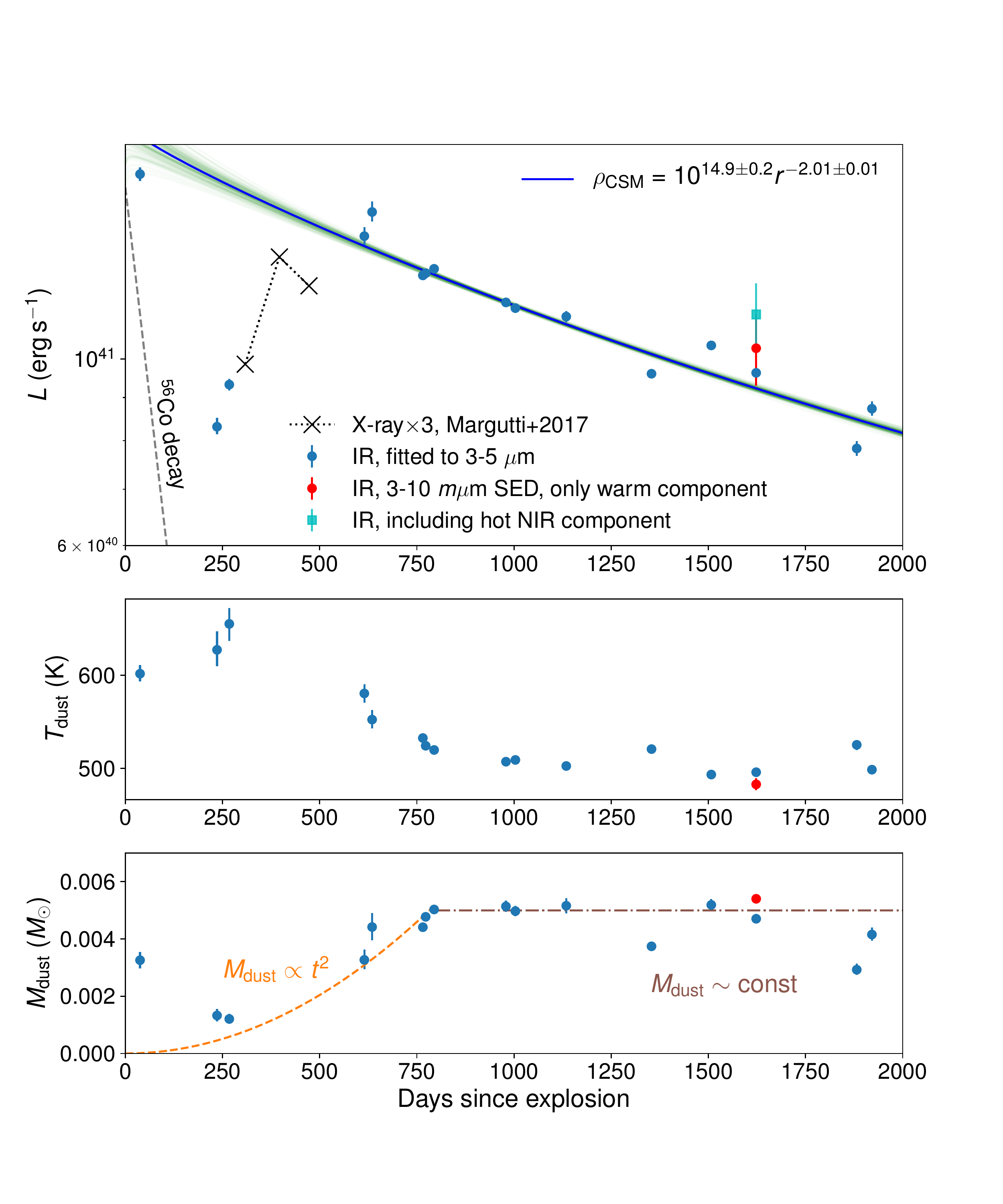}
    \caption{Dust luminosity (top), temperature (middle), and mass (bottom) from fitting a dust model to \textit{Spitzer} photometry. 
    We assume that the dust composition remains at 31\% silicate, 69\% carbonaceous, as derived from the 1-10 $\mu$m SED from 1620 days (see Fig.~\ref{fig:midIR_SED}). 
    The red points in all panels show the dust parameters of the warm component from the SED fitting including the 9.7 $\mu$m observation for comparison.
    The square cyan point in the top panel shows the IR luminosity from both the hot and warm components from the SED fit.
    The top panel includes the typical 0.01 mag/day light curve expected from radioactive decay of $\rm ^{56}Co$, to demonstrate the long-lasting nature of SN\,2014C.
    X-ray light curve from \cite{margutti2017} (multiplied by 3) is also included to show the approximated epoch of peak interaction. 
    The bolometric light curve model from \cite{moriya2013}, fitted to our data later than 600 days, is shown in this panel with the fitted parameters in the legend.
    Transparent lines in the background of the top panel are 100 individual models from the MCMC sample to representing the range of the fitted parameters.
    In the bottom panel, we plotted the $M_{\rm dust} \propto t^2$ line expected for an interaction with a constant density shell and the constant $M_{\rm dust}$ line expected for a wind-like CSM. 
    }
    \label{fig:LTM}
\end{figure}

To track the evolution of the dust parameters, we fit the SED at other epochs with \textit{Spitzer}/IRAC data, assuming a fixed ratio between carbonaceous and silicate dust.
We do not fit for the hot, near-IR component due to the lack of temporal coverage and its small contribution ($<10\%$) to the IR luminosity.
We run the MCMC fitting routine on all epochs and find that the posterior distributions of dust luminosity, temperature, and mass are Gaussian.
Those parameters are shown in in the top, middle, and bottom panels of Fig.~\ref{fig:LTM}, and are listed in Table~\ref{tab:dustLTM}.
At 1620 days, the epoch for which we have the full SED, we compare the values obtained for the warm dust component by fitting the full SED to those obtained from fitting only 3-5 $\mu$m data. 
This is to demonstrate that the discrepancy is minimal. 
In the top panel, we also show the full IR luminosity including both the hot and warm dust components to demonstrate that the hot component contributes minimally to the total IR luminosity. 
As a result, we use the luminosity of the warm dust component as an estimate of the bolometric luminosity of the SN.

\begin{table}
\centering
\caption{Luminosity, temperature, and mass of dust in the warm component of SN\,2014C, assuming 38\% silicate}
\label{tab:dustLTM}
\begin{tabular}{rrrrrrr}
\toprule
Epoch & $L$ & $\sigma_L$ & $T$ & $\sigma_T$ & $M$ & $\sigma_M$ \\
(day) & \multicolumn{2}{c}{($10^{40} \, \rm erg\,s^{-1}$)}   & \multicolumn{2}{c}{(K)} &   \multicolumn{2}{c}{($10^{-3} \, $\msun{})}  \\   
\hline 
37 & 16.6 & 0.3 & 602 & 9 & 3.2 & 0.3 \\
235 & 8.3 & 0.2 & 628 & 18 & 1.3 & 0.2 \\
267 & 9.3 & 0.1 & 651 & 16 & 1.2 & 0.2 \\
614 & 14.0 & 0.3 & 579 & 9 & 3.3 & 0.3 \\
634 & 15.0 & 0.4 & 551 & 10 & 4.4 & 0.5 \\
765 & 12.6 & 0.1 & 532 & 2 & 4.4 & 0.1 \\
772 & 12.7 & 0.1 & 524 & 2 & 4.8 & 0.1 \\
794 & 12.8 & 0.1 & 519 & 1 & 5.0 & 0.1 \\
979 & 11.7 & 0.1 & 507 & 2 & 5.1 & 0.2 \\
1003 & 11.5 & 0.1 & 509 & 2 & 4.9 & 0.2 \\
1134 & 11.2 & 0.2 & 502 & 3 & 5.1 & 0.2 \\
1353 & 9.6 & 0.1 & 520 & 2 & 3.7 & 0.1 \\
1507 & 10.4 & 0.1 & 493 & 2 & 5.2 & 0.2 \\
1623 & 9.6 & 0.1 & 496 & 2 & 4.7 & 0.2 \\
1881 & 7.8 & 0.1 & 525 & 5 & 2.9 & 0.2 \\
1920 & 8.7 & 0.2 & 499 & 4 & 4.1 & 0.2 \\
\hline
\end{tabular}
\end{table}

\subsection{Light Curve Modeling}\label{sec:LC_model}


In this section, we fit the light curve of SN\,2014C obtained in the last section with an analytic model from \cite{moriya2013}. 
The model assumes that the density profile in the SN ejecta follows a broken power law with $\rho_{\rm ej} \propto r^{-\delta, -n}$ for the inner and outer ejecta, respectively. 
These ejecta interact with the CSM, whose density profile is $\rho_{\rm CSM} = D r^{-s}$. 
The interaction starts in the outer ejecta, and then transitions into the inner ejecta at time $t_t$ post-explosion. 
Moriya et al. derived analytic expressions for the bolometric luminosity in the general CSM case where $s$ and $D$ are free parameters, for both $t < t_t$ (eq. 23) and $t > t_t$ (eq. 26).
The latter scenario is applicable to our data taken after 600 days, when the flux in the \textit{Spitzer} bands appear to dominate the SED.
We fit the following equation (eq. 26 in \citealp{moriya2013}) to our data
\begin{equation}
    L(t) = 2 \pi \epsilon D r_{\rm sh}(t)^{2-s} \left( \dfrac{(3-s) M_{\rm ej} \left( \frac{2 E_{\rm ej}}{M_{\rm ej}} \right)^{1/2}}{4 \pi D r_{\rm sh}(t)^{3-s} + (3-s) M_{\rm ej}} \right)^3
\end{equation}
where the time dependence is in $r_{\rm sh}(t)$, which is numerically solved using their eq. (12):
\begin{equation}
    \dfrac{4\pi D}{4-s} r_{\rm sh}(t)^{4-s} + (3-s) M_{\rm ej} r_{\rm sh}(t) - (3-s)M_{\rm ej} \left(\dfrac{2E_{\rm ej}}{M_{\rm ej}} \right)^{1/2} t = 0
\end{equation}
We will later compute $t_t$ to verify that our data satisfy $t > t_t$.

First, we perform an MCMC fit varying the CSM parameters: $D$ and $s$ from $\rho_{\rm CSM} = D r^{-s}$, and the explosion parameters: $M_{\rm ej}$ and $E_{\rm ej}$.
We impose minimal physical constraints: $M_{\rm ej}$, $E_{\rm ej}$, $D > 0$, and the explosion velocity $v_{\rm ej} = (2E_{\rm ej}/M_{\rm ej})^{0.5} < c$. 
The analytic model is only valid for $0 \leqslant s < 3$. 
We also try removing the degeneracy between $M_{\rm ej}$ and $E_{\rm ej}$ by imposing the observed photospheric velocity $v_{\rm phot} = (10E_{\rm ej}/3M_{\rm ej})^{0.5} = 13000 \, \rm km\,s^{-1}$ \citep{margutti2017, milisavljevic2015}. 
In both cases, MCMC fits do not converge as the explosion parameters mainly affect the shape of the light curve at early times.
The fitted explosion energy can grow arbitrarily large with a steep CSM density profile that makes very dense CSM close to the star. 

Instead, we use the explosion parameters inferred from early time observations: $M_{\rm ej} \sim 1.7$ \msun{} and $E_{\rm ej} \sim 1.8\times 10^{51}$ erg \citep{margutti2017}, and only fit for the CSM parameters.
The results we obtain, with 1-$\sigma$ uncertainty, are $D = 10^{14.9 \pm 0.2} \,(\rm cgs)$ and $s = 2.01 \pm 0.01$, which indicate that the CSM profile is wind-driven.
Recall that the parameter $D$ sets the absolute density scale of the CSM. 
In the wind driven scenario ($s=2$), $D = \dot{M}/(4 \pi v_w)$ where $\dot{M}$ and $v_w$ are the mass loss rate and the wind velocity, and $\dot{M}/v_w$ is dependent on the explosion parameters.
To derive the exact dependence, we solved eq. (35) and (36) in \cite{moriya2013} to eliminate $\epsilon$ and obtained $D  \propto \dot{M}/v_w \propto M_{\rm ej}^{3/2} \, E_{\rm ej}^{-1/2}$.   
From our fitted distribution for $D$ and this relation, we can write the mass loss rate as a function of $v_w$ and the explosion parameters as
\begin{equation}
    \dot{M} = (1.7\substack{+0.9\\-0.6})\times 10^{-3} \, \mathrm{{M_{\odot}}{yr^{-1}}} \, 
    \left(\frac{v_w}{100\, \rm{km\, s^{-1}}}\right)
    \left(\frac{M_{\rm ej}}{1.7\, M_{\odot}}\right)^{1.5}
    \left(\frac{E_{\rm ej}}{1.8 \times 10^{51} \, \rm{erg}}\right)^{-0.5}
\end{equation}
To check that we are in the $t > t_t$ limit, we compute the transition time using eq. (34) in \cite{moriya2013}, and found that $t_t = 345\pm 12$ days. 
Hence, our observations from $t > 600$ days are well beyond this limit. 

The density profile derived above allows us to estimate the total mass contained in the CSM.
With the wind-driven density profile $\rho \propto r^{-2}$, the gas mass contained up to radius $r$ is linearly proportional to $r$. 
The total (gas) mass of the wind-driven component of the CSM contained up to radius $r$ is simply $M_{\rm CSM} = 4 \pi D r$.
Since $D = 10^{14.9\pm 0.2} \, \rm g\, cm^{-1}$, the wind-driven portion of the CSM contains $M_{\rm CSM} = 5 \substack{+3\\-2} \times 10^{-18} \Delta r \, M_{\odot}/\rm cm $. 
As such, even if the outer CSM extends out to 10 times the current shock location, $10^{18}$ cm, the total wind-driven CSM mass is only about $ 5 \substack{+3\\-2}$ \msun{}.
More likely, the region of the CSM containing the dust we observe only extends out the order of its current size ($\sim 2 \times 10^{17}$ cm), with the total mass of about 1 \msun{}. 
Along with $\approx$1~\msun{} in the dense CSM shell \citep{margutti2017}, the total CSM mass around SN\,2014C is about $2 \substack{+0.6\\-0.4}$ \msun{}.
This total CSM mass constraint presents further issue for the pure carbonaceous dust scenario, which needs 0.22 \msun{} of carbonaceous dust to explain the 9.7 $\mu$m flux observed at 1620~d. 
Recall that this dust mass is a lower limit, assuming that dust is optically thin and not clumpy, of the total mass of dust emitting in the IR at that epoch. 
That quantity is itself a lower limit of the total mass of dust in the CSM since not all dust is illuminated by the shock interaction at that epoch. 
The very conservative upper limit of gas-to-dust ratio is then $9 \substack{+3 \\-2}$:1, which is at odds with dusty media observed around Galactic massive stars (e.g. $38\pm15$:1 in LBV WRAY 15-751 \citealp{Vamvatira-Nakou2013}) and SN remnants (e.g. the Crab Nebula has gas-to-dust ratio of 26-39:1, \citealp{owen2015}). 

\subsection{Evolution of the He 1.083 $\mu$m Line Profile}\label{sec:He_profile}

\begin{figure}
    \centering
    \includegraphics[width = \linewidth]{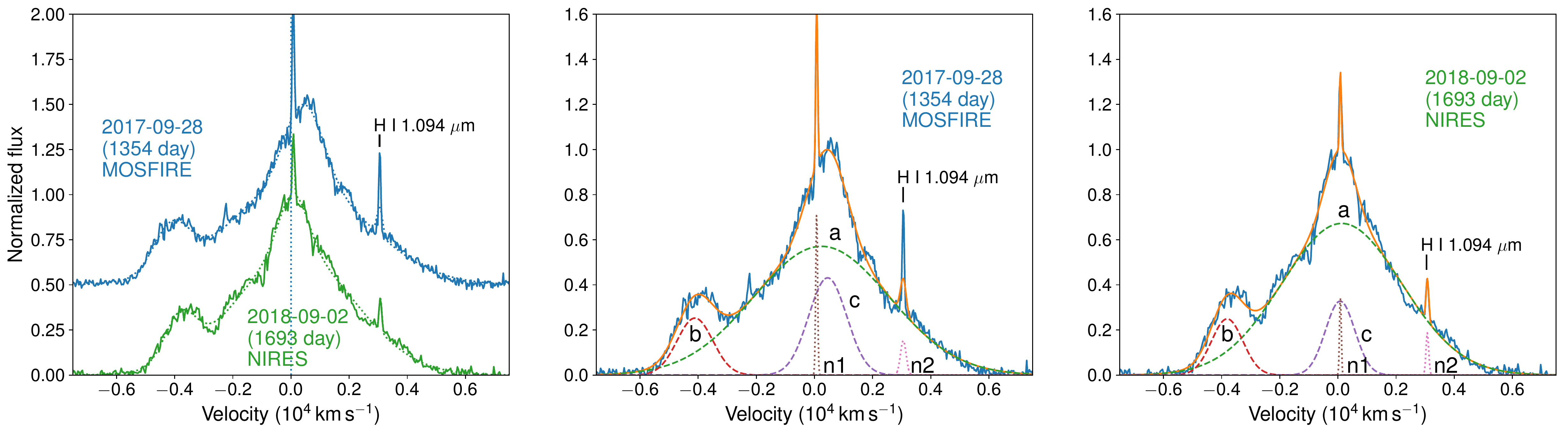}
    \caption{Left: The He \textsc{I} 1.083 $\mu$m line profile from our high-SNR spectra. Middle and right: line profiles from days 1354 and 1693 respectively. 
    Spectra are normalized so the peak of the broad component is 1. 
    For each epoch, we showed different Gaussian components used to fit the profile.
    For both epochs, the He 1.083 $\mu$m can be fit with one broad component (a, $2000 \,\rm km\,s^{-1}$), two intermediate width (b and c, $500\,\rm km\,s^{-1}$) components centering at 0 and 4000 $\rm km\,s^{-1}$, and lastly one narrow component (n1, $< 100 \rm km\,s^{-1}$) at rest velocity. 
    The n2 component is a narrow hydrogen 1.094 $\mu$m line.
    }
    \label{fig:He_profile}
\end{figure}

Fig.~\ref{fig:He_profile} (left) shows the evolution of the line profile of the broad He \textsc{I} 1.083 $\mu$m line.
The 2.058 $\mu$m line has a similar profile at all epochs, but the SNR is low as it is a much weaker line. 
We analyze spectra from days 1354 and 1693 because they have the highest SNR, and the other epochs that cover the $Y$ band (1305 and 1621 d) are close in time to these two epochs.
We model the profile of the 1.083 $\mu$m line with a combination of Gaussian components, and find that both epochs are well fit by four Gaussian components. 
Fig.~\ref{fig:He_profile} (middle and right) shows the fitting results. 
There is a broad component approximately at rest, marked as ``a", with a FWHM velocity of $\approx 2000 \, \rm km\,s^{-1}$.
This component is likely from the He-rich ejecta of the SN.
Second, there are two intermediate components, marked ``b" and ``c", at the central velocities of -4000 and about 0 $\rm km\,s^{-1}$. 
They are both with a common FWHM of 500 $\rm km\,s^{-1}$, suggesting that these intermediate lines are from the shocked CSM.
Lastly, one narrow component (n1) with FWHM $<100 \rm km\,s^{-1}$ is present.
The n2 component is the hydrogen 1.094 $\mu$m line.
The FWHM of all resolved lines (not n1 and n2) decreases by about 10\% between the two epochs. 
For comparison, the CSM interaction model used in \textsection\ref{sec:LC_model} also tracks the shock velocity. 
In the wind-driven case, we can differentiate eq. (19) in \cite{moriya2013} with respect to $t$ and get the shock velocity
\begin{equation}
    v_{\rm sh}(t) = \sqrt{\dfrac{2 E_{\rm ej}/M_{\rm ej}}{1+2at}}
    \quad\mathrm{where}\quad 
    a = \sqrt{\dfrac{2 E_{\rm ej}}{M_{\rm ej}^3}} \dfrac{\dot{M}}{v_w}
\end{equation}
According to this equation, the shock velocity would decrease by 5\% between 1354 and 1693 days, given the assumed ejecta mass, energy, and the fitted mass loss rate and wind velocity inferred from the light curve. 
This is roughly consistent with the deceleration observed in the ejecta. 
The ``b" component of the spectral profile may arise from the same hot spot found in the VLBI imaging \citep{bietenholz2018}. 

\section{Discussion and Conclusions}\label{sec:discussion}
\subsection{Multi-component CSM and potential binary origin}
The CSM observed in SN\,2014C clearly has several distinct components, both in density and composition.
As shown already in the literature \citep{milisavljevic2015, margutti2017, anderson2017}, the CSM has at least two density components: a low density bubble up to $\sim 10^{16} \, \rm cm$, and a dense, hydrogen-rich shell starting at $\sim 5 \times 10^{16} \, \rm cm$.
Continued IR observations, presented in this work, show that the CSM extends out to at least $1.4 \times 10^{17} \, \rm cm$ based on the current shock location at the latest \textit{Spitzer} epoch at 1920 d. 
The light curve shape, fitted with a CSM interaction model, further shows that the density profile of the outer part of the CSM differs from the constant density shell observed at early times.
The outer CSM density profile is well-explained by a steady wind-driven scenario ($\rho_{\rm CSM} \propto r^{-2}$) with a high mass loss rate of $\dot{M} \approx 10^{-3}$ \msun{}\,$\rm yr^{-1}$. 
The evolution of the observed dust mass also shows two distinct density profiles in the CSM. 
Before 750 days post-explosion, the observed dust mass is growing as $M_{\rm dust} \propto t^2$, indicating a CSM shell with constant density. 
After 750 days, the observed dust mass remain constant, indicating the CSM with a wind-driven ($\rho_{\rm CSM} \propto r^{-2}$) density profile. 
In addition to the distinct density components, we infer from the SED fitting that the CSM likely contains both carbonaceous and silicate grains, which form in different environments (C/O ratio $>$ 1 and $<1$, respectively), suggesting that the CSM was formed by stellar outflows with different compositions. 

One possible explanation for the density and chemical inhomogeneity is that the CSM is created by a progenitor that is transitioning from the oxygen rich phase (e.g. RSG or LBV) to the carbon rich (e.g. WC) phase, similar to what suggested by \cite{milisavljevic2015}.
In this scenario, the dense shell is formed by the fast WR wind sweeping up the slower RSG (or LBV) wind, which can mix in carbonaceous dust.  
There are three caveats of this scenario.
(i) The WR phase would only last for $\sim$ 10 yr since the wind velocity is $\sim 1000 \, \rm km\,s^{-1}$ and the CSM shell is at $\sim 10^{16} \, \rm cm$. This is much shorter than the expected life time of a WR.
Furthermore, the LBV (or RSG phase) are not expected to transition directly into the WC phase, but have to go through the WN phase first, making this scenario even less likely.
This caveat was mentioned in \cite{milisavljevic2015}.
We note, however, that recent observational studies have identified dusty WC systems, WR112 and Apep, with stagnant or slower expanding circumstellar dust ($< 600$ km s$^{-1}$; \citealp{lau2017,callingham2019}).
(ii) A single WR progenitor is incompatible with the low ejecta mass of $\sim$ 1.7 \msun{} inferred by \cite{margutti2017}.
(iii) The wind parameter of $\dot{M}/v_w =  1.7\pm 0.1 \times 10^{-3} \, M_{\odot} \, \mathrm{yr^{-1}}/100 \rm \, km\, s^{-1}$, inferred from our light curve fitting is inconsistent with the LBV quiescent wind phase. 
It is more consistent with either the high end of extreme RSG mass loss or the binary Roche-lobe overflow (RLOF) scenario \citep{smith2017}.

These caveats can be resolved by considering a binary progenitor system to SN\,2014C. 
In this scenario, the H-rich envelope of the progenitor star is stripped by binary interaction, not by its line-driven stellar wind \citep[e.g.][]{podsiadlowski1992}. 
As a result, the progenitor star need not be massive to be able to shed its hydrogen envelope, which is consistent with the ejected mass inferred from observations. 
\cite{margutti2017} discussed this scenario in details as one of the possible origins of the dense CSM shell around SN\,2014C (another scenario being late-stage nuclear burning instability). 
They found from binary population modeling that 3-10\% of SNe Ib/c progenitor went through a common-envelope mass loss scenario in which the stripped H-rich envelope can remain around the progenitor system long enough to interact with the SN. 
Such a binary progenitor system can also explain the chemical inhomogeneity and the density profile in the outer CSM. 
The density profile created by a binary system, assuming no interactions between two winds, is $\rho_{\rm CSM}(r) = (\dot{M}_1/v_1 + \dot{M}_2/v_2)/(4\pi r^2)$, where $\dot{M}_i$ and $v_i$ are the mass loss rate and wind velocity of the two components, and $r$ is much larger than the binary separation.
This is consistent with the observed density profile presented in this work. 
This scenario is also in line with the growing consensus that most stripped-envelope SNe are created via binary interaction, as opposed to single star mass loss \citep[e.g.][]{smith2014, demarco2017}. 
SN\,2014C stands out from the rest of SNe Ib/c perhaps due to the progenitor's binary parameter that caused the mass ejection event to occur right before core-collapse of one of the components. 

\subsection{RY Scuti as a Galactic analogue to the progenitor system of SN\,2014C}
There is a potential Galactic analogue to the progenitor system of SN\,2014C: the massive binary system RY Scuti \citep[][]{hjellming1973, gehrz1995,smith1999, gehrz2001, smith2002, smith2011b}.
RY Scuti is an interacting binary caught shortly after major mass transfer events.
It consists of a primary, mass-doner, O9/B0 supergiant star with a current mass of 8 \msun{} and a secondary, mass-gainer, O5 star with the current mass of 30 \msun{}. 
The initial masses for the primary and secondary stars were estimated to be about 20-25 and 15-20 \msun{}.
The He core mass of the primary star would have been roughly 6-8 \msun{} \citep{woosley1986}, indicating that it had lost most of its hydrogen envelope, and would soon become a WR star. 
High resolution imaging in the optical with \textit{Hubble Space Telescope (HST)} and in the infrared using adaptive optics (AO) revealed detached, expanding tori of CSM ejected in separate eruptive mass loss events \citep{smith1999, smith2002, smith2011b}.
The inner gas-rich torus was $3.4\times 10^{16}$ cm from the star, and was detected primarily in the optical.
The outer dusty torus was $5.4\times 10^{16}$ cm from the star, and was detected primarily in the IR.
Proper motion measurements from multi-epoch observations 6 (12) years apart in the infrared (optical) showed that the radial expansion velocity of the gas and dust tori was 40-50 $\rm km\, s^{-1}$.
The two tori had different velocities, and were ejected separately, about 100 yr apart \citep{smith2011b}.

The SN from the primary star of RY Scuti may resemble SN\,2014C in the following aspects.
First, the SN will initially be either of Type IIb or Ib/c, depending on how much hydrogen envelope is left. 
The SN will transition into a SN\,2014C-like event once the SN shock reaches the location of the dusty tori.
The strength of the interaction, however, may be much lower depending on the mass in the CSM. 
The outer dusty torus around RY Scuti was $5.4\times 10^{16}$ cm away from the star (measured from imaging), which is comparable to the location of the dense CSM shell around SN\,2014C \citep{margutti2017}. 
Hence, the timescale at which the shock interaction commences would be similar to that observed in SN\,2014C (about 100 days post-explosion). 
Second, the toroidal geometry of the CSM will allow the SN shock to propagate unimpeded, explaining the constant shock velocity despite CSM interaction \citep{bietenholz2018}. 
Third, the IR SED of the outer dusty torus observed by Keck Long Wave Spectrometer (LWS) showed a strong silicate dust feature at 10 $\mu$m, similar to what we inferred in SN\,2014C \citep{gehrz2001}. 
Moreover, \cite{gehrz2001} invoked a dust model with a mix of carbonaceous and astronomical silicate dust grains to explain the SED shape of RY Scuti, similar to what we did for SN\,2014C.
The temperature of the dust grains were slightly lower to what was observed in SN\,2014C (300-400 K as opposed to 500 K), which is not unexpected given the lower luminosity from  RY Scuti ($\approx 2 \times 10^{6} \, L_{\odot}$) as opposed to what would be expected in a SN explosion. 
While RY Scuti and SN\,2014C's progenitor may be similar in many aspects, they differ sharply in the CSM mass. 
SN\,2014C has of order 1 \msun{} in the CSM with about $5\times 10^{-3}$ \msun{} of dust; meanwhile, RY Scuti has only $1.4 \times 10^{-6}$ \msun{} of dust \citep{gehrz2001}. 
This led \cite{smith2011b} to conclude that the SN from RY Scuti's primary star will be an ordinary SN Ib/c and not an interacting SN IIn. 
They noted, however, that RY Scuti's mass transfer was very conservative, since about 15 \msun{} has been transferred to the secondary star with only about $0.003$ \msun{} lost into the CSM (gas mass of the inner torus, \citealp{smith2002}). 
The progenitor system to SN\,2014C may be a binary system much like RY Scuti, but with mass ratio and separation such that the mass transfer was much less conservative, resulting in much more mass being lost into the CSM. 

\subsection{Conclusion}
SN\,2014C's prolonged IR emission reveals two important characteristics of the CSM.
First, it likely contains a mixture of carbonaceous and silicate dust inferred from the SED including 9.7~$\mu$m photometry.
If confirmed, this is the first identification of silicate dust in an interacting SN. 
Alternatively, the SED could also be fitted with purely carbonaceous dust with a cold, 300 K, component containing 0.22 \msun{} of dust grains. 
That scenario would be unprecedented as well.
Second, the light curve fitting shows that the CSM extends out to at least $1.4 \times 10^{17}$~cm, with the wind-driven density profile in the outer part.
This component is in addition the the inner low density bubble and the dense shell with a constant density identified earlier in the literature.
The chemical inhomogeneity of the CSM suggests that the progenitor system of SN\,2014C is likely a binary with one component producing an oxygen-rich outflow and another component carbon-rich outflow.
While most SNe Ib/c are not observed to interact with their lost hydrogen envelope, SN\,2014C may present the extreme end of the population, where the binary-driven mass ejection event that strips the progenitor's envelope happens just before the core-collapse, leaving a dense, detached H-rich CSM to interact with the SN shock.
The progenitor system that produced SN\,2014C may look very similar to RY Scuti, but with much more mass ejected from the binary system into the CSM.
This work underscores the importance of late-time IR observations to constrain CSM properties of interacting SNe, since the outer part of the CSM can only be probed by late time interactions, which emit mostly in the IR. 
Assuming that this CSM does not abruptly end at this radius, the interaction will still be ongoing by the time that \textit{JWST} becomes operational. 
It will enable IR spectroscopic observations of this SN, and of other interacting SNe, which will help us understand SN\,2014C's place in the continuum of SN types from Ib/c to IIn, and to construct a complete picture of how mass loss operates at the end of the life of a massive star. 

\acknowledgments 
We thank Takashi Moriya, Itsuki Sakon, and Takashi Onaka for helpful discussions. 
We thank Takuya Fujiyoshi for conducting the COMICS observations.
This work was supported by the GROWTH (Global Relay of Observatories Watching Transients Happen) project funded by the National Science Foundation under PIRE grant No. 1545949. 
GROWTH is a collaborative project among California Institute of Technology (USA), University of Maryland College Park (USA), University of Wisconsin Milwaukee (USA), Texas Tech University (USA), San Diego State University (USA), University of Washington (USA), Los Alamos National Laboratory (USA), Tokyo Institute of Technology (Japan), National Central University (Taiwan), Indian Institute of Astrophysics (India), Indian Institute of Technology Bombay (India), Weizmann Institute of Science (Israel), The Oskar Klein Centre at Stockholm University (Sweden), Humboldt University (Germany), Liverpool John Moores University (UK), and University of Sydney (Australia).
KM acknowledges support by JSPS KAKENHI Grant (18H04585, 18H05223, 17H02864).
RDG is supported by NASA and the United States Air Force.
Some of the data presented herein were obtained at the W. M. Keck Observatory, which is operated as a scientific partnership among the California Institute of Technology, the University of California and the National Aeronautics and Space Administration. The Observatory was made possible by the generous financial support of the W. M. Keck Foundation.
This work is based in part on observations obtained at the Gemini Observatory, which is operated by the Association of Universities for Research in Astronomy, Inc., under a cooperative agreement with the NSF on behalf of the Gemini partnership: the National Science Foundation (United States), National Research Council (Canada), CONICYT (Chile), Ministerio de Ciencia, Tecnolog\'{i}a e Innovaci\'{o}n Productiva (Argentina), Minist\'{e}rio da Ci\^{e}ncia, Tecnologia e Inova\c{c}\~{a}o (Brazil), and Korea Astronomy and Space Science Institute (Republic of Korea).
This work is based in part on data collected at Subaru Telescope, which is operated by the National Astronomical Observatory of Japan.
The three aforementioned observatories are on the summit of Maunakea and the authors wish to recognize and acknowledge the very significant cultural role and reverence that the summit of Maunakea has always had within the indigenous Hawaiian community.  
We are most fortunate to have the opportunity to conduct observations from this mountain.
Some of the data presented herein were obtained at Palomar Observatory, which is operated by a collaboration between California Institute of Technology, Jet Propulsion Laboratory, Yale University, and National Astronomical Observatories of China. 
This work is based in part on observations made with the \textit{Spitzer Space Telescope}, which is operated by the Jet Propulsion Laboratory, California Institute of Technology under a contract with NASA. Support for this work was provided by NASA through an award issued by JPL/Caltech.
This research has made use of the NASA/IPAC Extragalactic Database (NED), which is operated by the Jet Propulsion Laboratory, California Institute of Technology, under contract with the National Aeronautics and Space Administration.
This research made use of Astropy, a community-developed core Python package for Astronomy \citep{astropy2018}.

\facilities{Hale (WIRC, TripleSpec), Spitzer, Keck (MOSFIRE, NIRES), Subaru (COMICS), Gemini (NIRI, GNIRS)}
\software{Astropy \citep{astropy2018}, Spextool \citep{cushing2004},MOSFIRE data reduction pipeline \citep{mclean2012}, xtellcor \citep{vacca2003}, emcee \cite{emcee}, Matplotlib \citep{Hunter:2007}, scipy \citep{scipy}}

\bibliography{main.bib}

\begin{appendix}
\counterwithin{figure}{section}

\section{MCMC fitting of the SED and Light Curve}
\subsection{SED fitting with Dust Models}\label{sec:MCMC_SED}
\begin{figure}
    \centering
    \includegraphics[width = 0.7\textwidth]{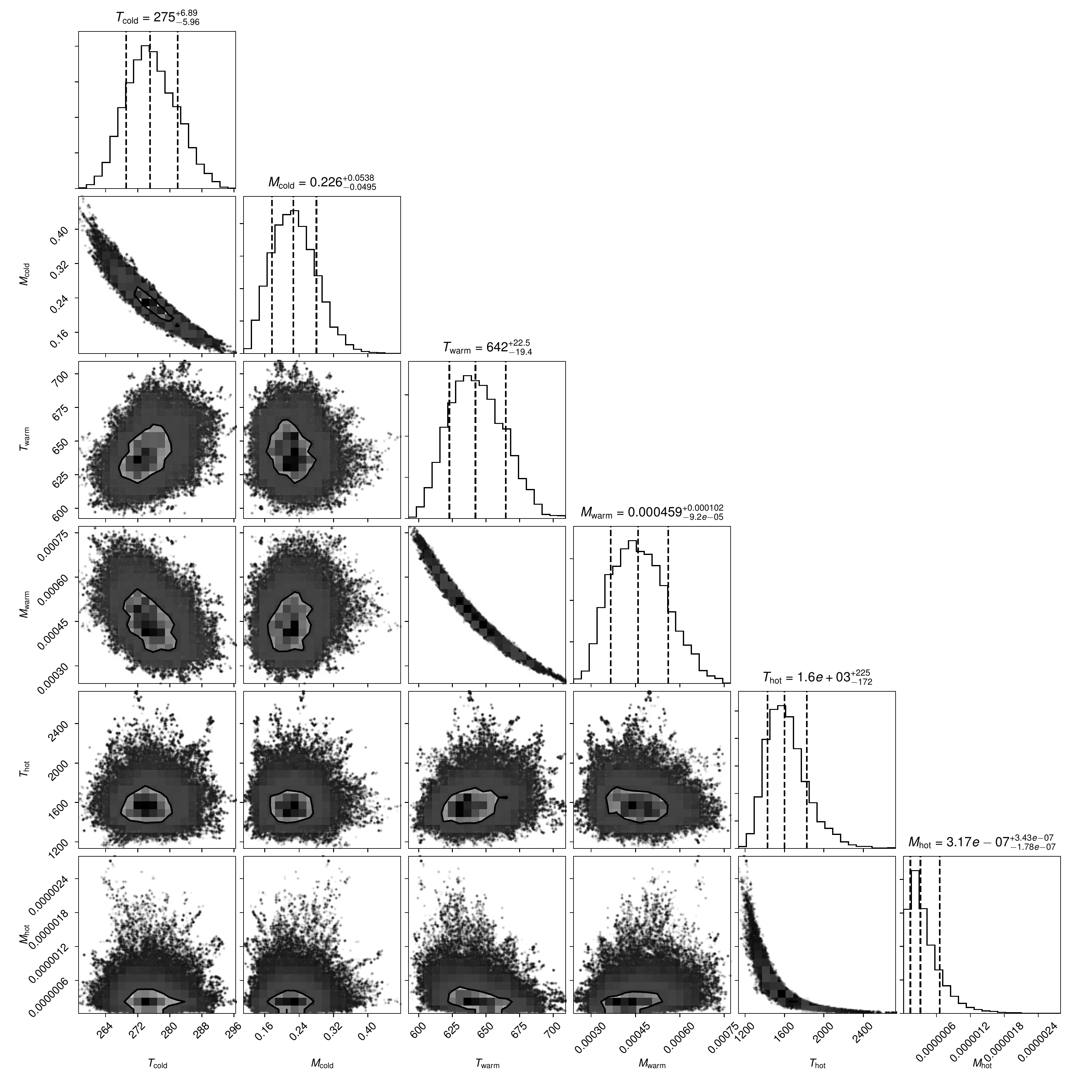}
    \caption{The corner plot showing the results of MCMC fitting of the SED shown in Fig.~\ref{fig:midIR_SED} using a dust model with 3 temperature components. 
    Each component has two parameters: temperature and mass. 
    All components only have carbonaceous dust with the grain size of 0.1 $\mu$m.
    Units of mass and temperatures are \msun{} and K respectively.  
    }
    \label{fig:corner_SED_C}
\end{figure}

To determine best-fit values for the dust parameters in the SED fittings in \textsection\ref{sec:si_dust}, and the CSM parameters in the light curve fitting in \textsection\ref{sec:LC_model}, we used the \texttt{emcee} package to run MCMC on our data. 
For the first model with all carbonaceous dust, there were 6 parameters: $M_{\rm cold, warm, hot}$ and $T_{\rm cold, warm, hot}$.
The dust composition was purely carbonaceous and the grain size was 0.1 $\mu$m. 
The initial values of these parameters were obtained by running a more traditional least square fit on the data using the \texttt{curve\_fit} routine in the \texttt{scipy.optimize} package. 
The prior distribution for all parameters are uniform, with only positive values allowed for mass and temperatures. 
We ran MCMC using 400 walkers for 1000 steps, and determined that the convergence happened after 200 steps.
Fig.~\ref{fig:corner_SED_C} shows the resulting posterior distribution of the 6 parameters, along with the median and $\pm 1\sigma$ values.

\begin{figure}
    \centering
    \includegraphics[width = 0.7\textwidth]{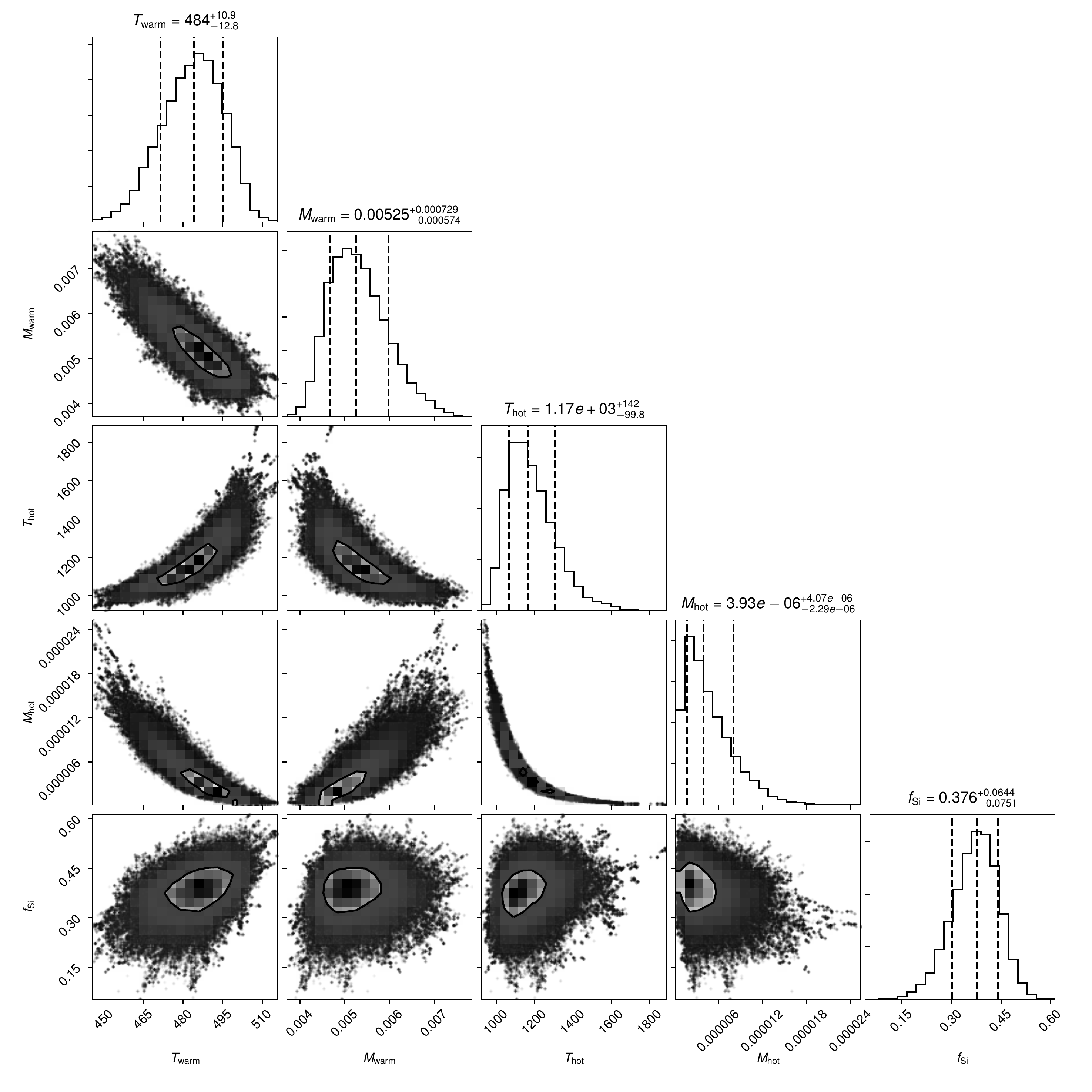}
    \caption{The corner plot showing the results of MCMC fitting of the SED shown in Fig.~\ref{fig:midIR_SED} using a dust model with two temperature components. The warm component has a mixture of carbonaceous and silicate dust with the mass fraction of $f_{\rm Si}$ in silicate. The total mass and temperature of this component are $M_{\rm warm}$ and $T_{\rm warm}$ respectively. The hot component only consists of carbonaceous dust, though both composition looks identical at this temperature. The total mass and temperature of this component are $M_{\rm hot}$ and $T_{\rm hot}$ respectively. Units of mass and temperatures are \msun{} and K respectively.  
    }
    \label{fig:corner_SED}
\end{figure}

For the second model with a mixture of carbonaceous and silicate dust, there were five free parameters: $M_{\rm warm}$, $T_{\rm warm}$, $M_{\rm hot}$, $T_{\rm hot}$, and $f_{\rm Si}$. 
The prior distribution was also uniform with positive mass and temperatures.
The silicate fraction $f_{\rm Si}$ could be between 0 and 1. 
The initial values were also obtained with a least square fit using \texttt{curve\_fit}.
We ran MCMC using 400 walkers for 2000 steps, and determined that the convergence happened after 1000 steps.
We note that the temperature and mass of the hot component are not as well constrained because the near-IR SED deviates from the dust model, potentially due to poor background subtraction. 
Fig.~\ref{fig:corner_SED} shows the resulting posterior distribution of the 5 parameters, along with the median and $\pm 1\sigma$ values.

\subsection{Light Curve Fitting with CSM Models}\label{sec:MCMC_LC}
\begin{figure}
    \centering
    \includegraphics[width = 0.45\textwidth]{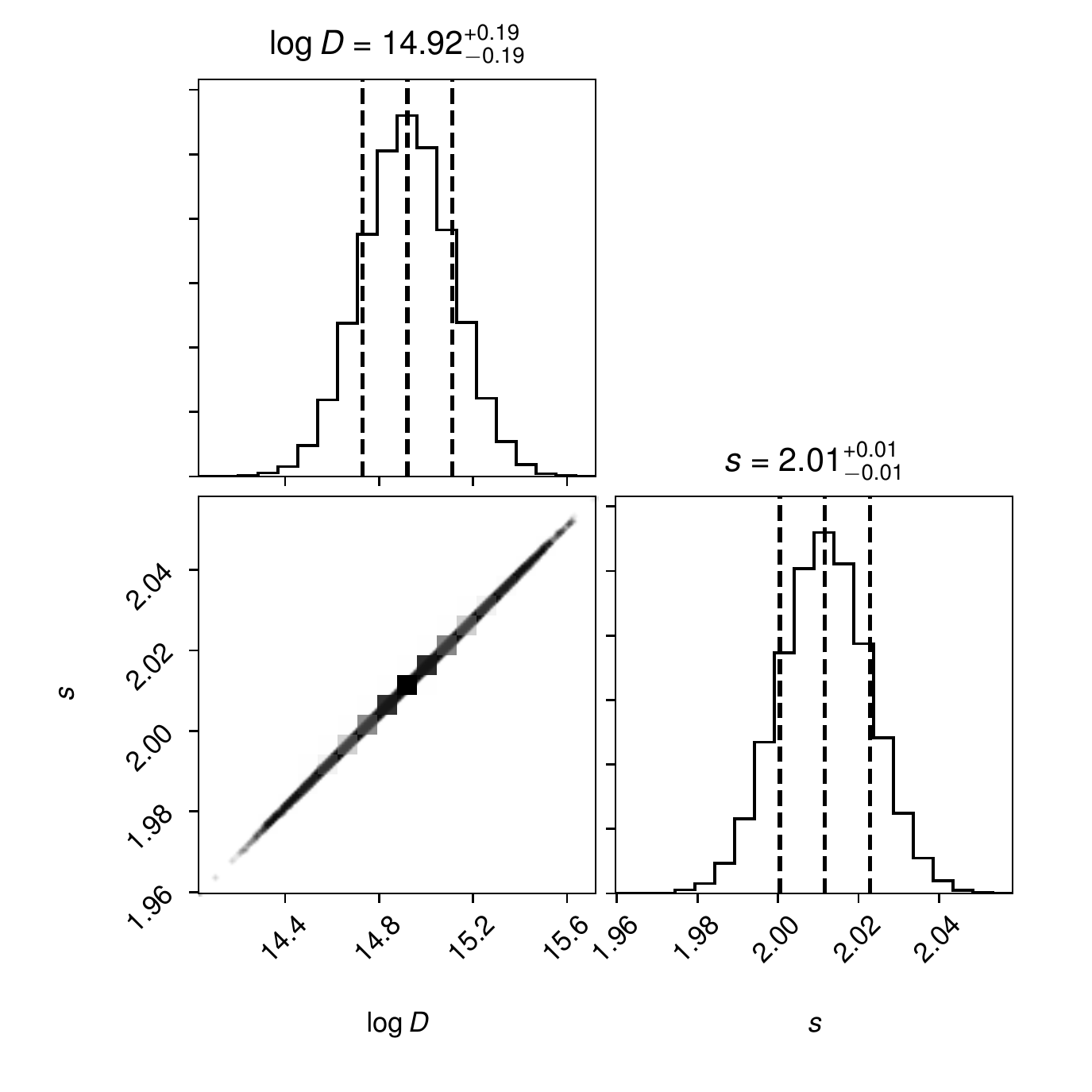} 
    \caption{The corner plot showing the results of MCMC fitting of the light curve from \textsection\ref{sec:LC_model} where we fitted a light curve of the SN interacting with a CSM with the density profile $\rho_{\rm CSM} = D r^{-s}$ with $D$ and $s$ as the free parameters in the fit.  
    }
    \label{fig:corner_LC}
\end{figure}

As discussed in the main text, we found that the MCMC fit can only constrain CSM density parameters, namely $D$ and $s$ from $\rho_{\rm CSM} = D r^{-s}$, and not the explosion parameters ($E_{\rm ej}$ and $M_{\rm ej}$). 
We considered a model with $D$ and $s$ as free parameters.
Similarly to the SED fit, the initial values were determined using least square fitting. 
We also used a uniform prior distribution with physical parameters set to positive numbers, and $0 \leqslant s < 3$.
Fig.~\ref{fig:corner_LC} shows the corner plots for the fit. 

\end{appendix}

\end{document}